\documentclass[11pt]{article}
\pdfoutput=1
\usepackage{color}
\usepackage[textwidth = 430 pt, textheight = 630 pt]{geometry}
\definecolor{MyDarkBlue}{rgb}{0.15,0.25,0.45}
\usepackage{epsfig,rotating}
\usepackage{amsmath,amsthm,amssymb}
\usepackage{amsfonts}
\usepackage{mathrsfs}
\usepackage{bbm}
\usepackage{bm}
\usepackage{hyperref}
\hypersetup{
colorlinks=true,
citecolor=MyDarkBlue,
linkcolor=MyDarkBlue,
urlcolor=MyDarkBlue,
pdfauthor={Derek Harland, Sam Palmer and Christian S\"amann},
pdftitle={Magnetic Domains},
pdfsubject={hep-th}
breaklinks=true
}

\let\fn\footnote
\renewcommand{\footnote}[1]{\linespread{1.1}\fn{#1}\linespread{1.29}}

\makeatletter\renewcommand{\section}{\@startsection
{section}{1}{\z@}{-3.5ex plus -1ex minus
    -.2ex}{2.3ex plus .2ex}{\bf }}
\makeatletter\renewcommand{\subsection}{\@startsection{subsection}{2}{\z@}{-3.25ex
plus -1ex minus
   -.2ex}{1.5ex plus .2ex}{\it }}
\makeatletter\renewcommand{\subsubsection}{\@startsection{subsubsection}{3}{-2.45ex}{-3.25ex
plus -1ex minus -.2ex}{1.5ex plus .2ex}{\it }}
\renewcommand{\thesection}{\arabic{section}}
\renewcommand{\thesubsection}{\arabic{section}.\arabic{subsection}}
\renewcommand{\@seccntformat}[1]{\@nameuse{the#1}.~~}

\renewcommand{\theequation}{\thesection.\arabic{equation}}
\makeatletter \@addtoreset{equation}{section}

\renewenvironment{thebibliography}[1]
     {\baselineskip=16pt plus 2pt minus 1pt
      \section*{\large\refname
        \@mkboth{\MakeUppercase\refname}{\MakeUppercase\refname}}%
     \list{\@biblabel{\@arabic\c@enumiv}}%
           {\settowidth\labelwidth{\@biblabel{#1}}%
            \leftmargin\labelwidth
            \advance\leftmargin\labelsep
            \@openbib@code
            \usecounter{enumiv}%
            \let\p@enumiv\@empty
            \renewcommand\theenumiv{\@arabic\c@enumiv}}%
      \sloppy
      \clubpenalty4000
      \@clubpenalty \clubpenalty
      \widowpenalty4000%
      \sfcode`\.\@m}

\newcommand{\appendices}{
\section*{Appendix}\label{appendices}\setcounter{subsection}{0}
\addcontentsline{toc}{section}{Appendix}
\setcounter{equation}{0}
\makeatletter
\renewcommand{\theequation}{\Alph{subsection}.\arabic{equation}}
\renewcommand{\thesubsection}{\Alph{subsection}}
\@addtoreset{equation}{subsection}
\makeatother
}

\theoremstyle{plain} 
\newtheorem{thm}{Theorem}

\newtheorem{conj}{Conjecture}

\def\periodb#1{\setbox0=\hbox{$#1$}#1\hskip-\wd0\hbox to\wd0{-}}

\newcommand{\id}{\mathrm{id}}   			

\newcommand{\CA}{\mathcal{A}}    			

\newcommand{\CC}{\mathcal{C}}

\newcommand{\CF}{\mathcal{F}}

\newcommand{\CH}{\mathcal{H}}
\newcommand{\CCH}{\mathscr{H}}
\newcommand{\CI}{\mathcal{I}}

\newcommand{\CL}{\mathcal{L}}

\newcommand{\CN}{\mathcal{N}}
\newcommand{\CO}{\mathcal{O}}

\newcommand{\CT}{\mathcal{T}}
\newcommand{\CCT}{\mathscr{T}}

\newcommand{\frg}{\mathfrak{g}}				

\newcommand{\FR}{\mathbbm{R}}     			
\newcommand{\FC}{\mathbbm{C}}     			
\newcommand{\NN}{\mathbbm{N}}     			
\newcommand{\RZ}{\mathbbm{Z}}     			
\newcommand{\CPP}{{\mathbbm{C}P}}    			

\newcommand{\trng}{t_\circ}
\newcommand{\trngd}{\dot{t}_\circ}
\newcommand{\urng}{u_\circ}

\newcommand{\dd}{\mathrm{d}}     			
\newcommand{\dpar}{\partial}     			
\newcommand{\embd}{{\hookrightarrow}}     		
\newcommand{\de}{\mathrm{e}}     			
\newcommand{\di}{\mathrm{i}}     			
\newcommand{\eps}{{\varepsilon}}			

\newcommand{\zb}{{\bar{z}}}

\newcommand{\eand}{{~~~\mbox{and}~~~}}     		
\newcommand{\ewith}{{~~~\mbox{with}~~~}}

\newcommand{\der}[1]{\frac{\dpar}{\dpar #1}}   		
\newcommand{\dder}[1]{\frac{\dd}{\dd #1}}   		
\newcommand{\delder}[1]{\frac{\delta}{\delta #1}}   		
\newcommand{\derr}[2]{\frac{\dpar #1}{\dpar #2}}   	
\newcommand{\dderr}[2]{\frac{\dd #1}{\dd #2}}   	
\newcommand{\tr}{\,\mathrm{tr}\,}     			

\newcommand{\agl}{\mathfrak{gl}}     			

\newcommand{\au}{\mathfrak{u}}
\newcommand{\asu}{\mathfrak{su}}
\newcommand{\aso}{\mathfrak{so}}

\newcommand{\sU}{\mathsf{U}}     			

\newcommand{\sSU}{\mathsf{SU}}

\newcommand{\sMat}{\mathsf{Mat}}

\newcommand{\sSO}{\mathsf{SO}}

\newcommand{\sEnd}{\mathsf{End}\,}

\newcommand{\acton}{\vartriangleright}     			
\newcommand{\remark}[1]{}     				
\def\tyng(#1){\hbox{\tiny$\yng(#1)$}}			
\def\tyoung(#1){\hbox{\tiny$\young(#1)$}}			

\newcommand{\pa}{\partial}
\newcommand{\yd}{\dot{y}}

\begin{document}

\begin{titlepage}
\begin{flushright}
 HWM--12--04 \\ EMPG--12--08
\end{flushright}
\vskip 2.0cm
\begin{center}
{\LARGE \bf Magnetic Domains}
\vskip 1.5cm
{\Large Derek Harland$^a$, Sam Palmer$^b$ and Christian S\"amann$^b$}
\setcounter{footnote}{0}
\renewcommand{\thefootnote}{\arabic{thefootnote}}
\vskip 1cm

{\em ${}^a$Department of Mathematical Sciences\\
Loughborough University\\
Loughborough, Leics., LE11 3TU, UK\\[0.5cm]
 
${}^b$Department of Mathematics and \\
Maxwell Institute for Mathematical Sciences\\[0.1cm]
Heriot-Watt University\\
Colin Maclaurin Building, Riccarton, Edinburgh EH14 4AS, U.K.}\\[0.5cm]
{Email: {\ttfamily d.g.harland@lboro.ac.uk~,~sap2@hw.ac.uk~,~c.saemann@hw.ac.uk}}
\end{center}

\vskip 1.0cm
\begin{center}
{\bf Abstract}
\end{center}
\begin{quote}
Recently a Nahm transform has been discovered for magnetic bags, which are conjectured to arise in the large $n$ limit of magnetic monopoles of charge $n$.  We interpret these ideas using string theory and present evidence for this conjecture.  Our main result concerns the extension of the notion of bags and their Nahm transform to higher gauge theories and arbitrary domains.  Bags in four dimensions conjecturally describe the large $n$ limit of $n$ self-dual strings.  We show that the corresponding Basu-Harvey equation is the large $n$ limit of an equation describing $n$ M2-branes, and that it has a natural interpretation in loop space.  We also formulate our Nahm equations using strong homotopy Lie algebras.
\end{quote}

\end{titlepage}

\tableofcontents

\newpage

\section{Introduction and results}

When D1-branes are stretched between two parallel D3-branes, magnetic monopoles are induced on the worldvolume of the D3-branes \cite{Diaconescu:1996rk,Tsimpis:1998zh}. Such configurations are described equally well either by the transverse fluctuations of the D1-branes or by the profile and curvature of the D3-branes. The former consist of solutions to the Nahm equation, and the latter are solutions of the Bogomolny equation in the non-abelian gauge theory. The Nahm transform \cite{Nahm:1981nb,Hitchin:1983ay} switches between these two descriptions.

In recent years there have been attempts to generalize the Nahm transform to describe M2-branes stretched between M5-branes.  In the case of one M5-brane and an arbitrary number of M2-branes, the analog of the Bogomolny equation is known as the self-dual string equation \cite{Howe:1997ue}.  The analog of the Nahm equation for one or two M2-branes is known as the Basu-Harvey equation \cite{Basu:2004ed}, and a generalization to arbitrarily many M2-branes was proposed in \cite{Terashima:2008sy,Gomis:2008vc,Hanaki:2008cu}.  There is currently no universal agreement on how the Nahm transform for M-branes works.

The extension of the Nahm transform to certain configurations of infinitely many D1-branes was developed in  \cite{Harland:2011tm}. The crucial observation is that the Lie algebra $\au(n)$ can be viewed as the algebra of functions on the fuzzy sphere, with $1/n$ playing the role of the non-commutativity parameter. The fields describing the transverse fluctuations of the D1-brane are $\au(n)$-valued functions on an interval $\CI$. In the limit $n\rightarrow\infty$ they become functions on $S^2 \times \CI$. These fields are then put together into a map
\begin{equation}
t:S^2\times \CI \rightarrow\FR^3~,
\end{equation}
from which the fields on $\FR^3$ can easily be constructed.

The resulting configurations are known as magnetic bags.  Magnetic bags are abelian configurations that were introduced in \cite{Bolognesi:2005rk}.
They are widely believed to describe the large $n$ limit of $n$-monopoles in non-abelian gauge theory. This is known as Bolognesi's conjecture. 

In the present article we investigate various extensions of the Nahm transform, in particular also to bags of self-dual strings. We begin in section \ref{sec:Bags} with a discussion of the 3-dimensional situation.  The notion of magnetic bags is generalized to that of {\em magnetic domains}; the latter may appear as limits not only of monopoles, but also of monopole walls, monopole chains, and probably other configurations.  We state and prove a Nahm transform for magnetic domains which generalizes that given in \cite{Harland:2011tm}.  We also give a partial proof of Bolognesi's conjecture for the case of magnetic discs, which are flattened magnetic bags.

In section \ref{sec:branes} we present a D-brane interpretation of magnetic bags and their Nahm transform.  The surfaces of magnetic bags are junctions of D-branes which are related by T- and S-duality to junctions of $(p,q)$ 5-branes.  These junctions appear in the Nahm data as defects.  The D-brane picture is valuable not only as further support for the magnetic bag conjecture, but also as a guide in generalizing the magnetic bag conjecture to M-theory.  Indeed, it seems very likely that $n$ M2-branes stretching between two M5-branes will form a bag as $n\to\infty$.  A striking feature here is that the bags are abelian, and thus evade the usual difficulties associated with writing down non-abelian higher gauge theories.

In sections \ref{sect:SDS}--\ref{sect:LSSDS} we investigate in detail bags and more general domains formed by self-dual strings.  A precise definition of these domains is formulated in section \ref{sect:SDS}, and we state and prove the Nahm transform for them.  The Nahm-dual picture for a self-dual string bag consists of solutions of the Basu-Harvey equation based on the algebra of functions on the 3-sphere.  These can be combined into a map
\begin{equation}
t:S^3\times \CI\rightarrow\FR^4~,
\end{equation}
from which the bag can be recovered.  This substantially improves a result of Ho and Matsuo \cite{Ho:2008nn}, who showed that the Bagger-Lambert-Gustavsson action based on the algebra of functions on a 3-manifold at least has the correct low energy degrees of freedom to describe M5-branes.

We go on to show in section \ref{sect:HermitianBH} that this Basu-Harvey equation is the large $n$ limit of the equation introduced in \cite{Terashima:2008sy,Gomis:2008vc,Hanaki:2008cu} for describing $n$ M2-branes.  The bags obtained in this large $n$ limit are quite constrained: they are necessarily invariant under a certain action of $\sU(1)$.  In fact, they can be identified with ordinary magnetic bags using the Hopf fibration.

Recently the idea has emerged that self-dual strings and their Nahm transform can be described using loop space \cite{Gustavsson:2008dy,Saemann:2010cp,Palmer:2011vx}.  We show in section \ref{sect:LSSDS} that our Nahm equation for self-dual string bags also has a natural loop space formulation and re-interpret the Nahm transform from that perspective.

Finally, we provide in section \ref{sec:higher} a construction for bags in higher gauge theories.  An interesting feature here is that the Nahm equation can be written as a Maurer-Cartan equation for an element of an $L_\infty$-algebra.

We hope that the results presented here will be of some use in obtaining a better understanding of self-dual strings.  Certainly, self-dual string bags are easier to write down than self-dual strings with charge $n<\infty$.  The Nahm transform for bags is very transparent, and should provide a consistency check on putative Nahm transforms for self-dual strings.

\section{Magnetic domains in three dimensions}\label{sec:Bags}

\subsection{From magnetic monopoles to magnetic domains}

In $\sSU(2)$ Yang-Mills-Higgs theory, we have an $\sSU(2)$ principal bundle over $\FR^3$ with connection 1-form $A$, curvature 2-form $F$ and an adjoint Higgs field $\Phi$. We define
\begin{equation}
F=\dd A +e A\wedge A~,~~~\dd_A\Phi=\dd\Phi+e[A,\Phi]~,
\end{equation}
where $e$ is the Yang-Mills coupling constant. The Yang-Mills-Higgs energy functional 
\begin{equation}
E = \tfrac{1}{2}\int_{\FR^3} \tr\left(F\wedge\ast F + \dd_A\Phi\wedge * \dd_A \Phi\right)
\end{equation}
admits a Bogomolny bound
\begin{equation}
E = \int_{\FR^3} \tr\left(\tfrac{1}{2}|\dd_A\Phi-*F|^2+\dd_A\Phi\wedge F\right) \ge \int_{S^2_\infty} \tr(F\Phi)~.
\end{equation}

This bound is saturated (and the Yang-Mills-Higgs equations of motion are satisfied) for BPS monopoles, which are defined as solutions $(A,\Phi)$ to the Bogomolny monopole equation 
\begin{equation}\label{eq:BogomolnyMonopole}
F=*\dd_A\Phi~,
\end{equation}
together with the asymptotic condition $||\Phi||:=\sqrt{\frac{1}{2}\tr(\Phi^\dagger\Phi)}\rightarrow v>0$ as $r\rightarrow\infty$.  This asymptotic condition on $\Phi$ breaks the gauge symmetry to $\sU(1)$, and therefore it makes sense to talk about the magnetic charge $q$ of a monopole.  It is well-known that the magnetic charge is quantized:
\begin{equation}\label{magnetic charge}
q := -\tfrac{1}{2}\int_{S^2_\infty} \frac{\tr(F\Phi)}{\|\Phi\|} = \frac{2\pi n}{e}~.
\end{equation}
Here $n\in \RZ$ is a topological charge which counts the number of monopoles. The Bogomolny bound can now be written as $E\ge vq$.

In this paper, we are interested in monopole configurations that arise in the limit $n\rightarrow \infty$.  For example, consider a BPS configuration of an odd number $n$ of monopoles in $\FR^3$ located on a one-dimensional lattice at $\vec{x}=(i,0,0)$, $i\in \RZ$, $|i|\leq (n-1)/2$, where we use the usual Cartesian coordinates on $\FR^3$.  Such configurations of monopoles are known to exist, and in a certain limit $n,v\to\infty$ one obtains a solution of the Bogomolny equation invariant under a translation group $\RZ$ \cite{Dunne:2005pr,Harland:2009yh}. This is an example of a {\em monopole chain} \cite{Cherkis:2000cj,Ward:0505254}.

Similarly, one can consider doubly-periodic monopoles invariant under the action of $\RZ^2$, given by $(x^1,x^2,x^3)\mapsto(x^1+i,x^2+j,x^3)$ for $i,j\in\RZ^2$.  One has the freedom to impose different boundary conditions as $z\to\pm\infty$, and configurations satisfying $\|\Phi\|\to A$ as $z\to-\infty$ and $\|\Phi\|\sim Bz$ as $z\to\infty$ for constants $A,B$ are know as {\em monopole walls}\footnote{Configurations for which $\|\Phi\|\sim B|z|$ as $z\to\pm\infty$ are called {\em monopole sheets}.} \cite{Lee:1998isa,Ward:0505254,Ward:2006wt}.  If a monopole wall has non-zero charge per unit period, then the total charge $n$ is again infinite.

Inductive reasoning might lead one to consider triply-periodic monopoles, but the following argument shows that there are no non-trivial examples of these.  Any triply-periodic monopole would correspond to a monopole on the compact manifold $T^3$.  The equation of motion $\square_A \Phi=0$ would then imply that $0=\int_{T^3}\tr(\Phi~\dd_A\ast \dd_A\Phi)=-\int_{T^3} \tr(\dd_A\Phi\wedge\ast \dd_A\Phi)$, and hence that $\dd_A\Phi$ vanishes\footnote{Note that although $\Phi$ is a section of an associated vector bundle, the expression under the integral is globally defined.}.

Our final examples of monopoles with $n\to\infty$ are {\em magnetic bags} \cite{Bolognesi:2005rk}.  Heuristically, a magnetic bag with finite charge $n$ consists of a finite-area segment of a monopole wall, folded around to form a closed surface.  The existence of such monopoles is an open question (which we discuss further in section \ref{commentsConjecture}), however, five examples are known with $n=3,4,5,7,11$ \cite{Lee:2008ze}.  These magnetic bags are roughly spherical in shape, and the lattice structures on their surfaces resemble the five Platonic solids.\footnote{There are actually two types of magnetic bag, termed ``abelian'' and ``non-abelian'' in \cite{Lee:2008ze}, but this distinction becomes irrelevant in the limit $n\to\infty$ that we consider.}  The size of the Platonic monopoles has been shown to be in good agreement with predictions of the bag model \cite{Manton:2011vm}.  Constructing further examples of magnetic bags on $\FR^3$ is difficult, because there are no further Platonic solids 
whose symmetries can be exploited.  The situation is much better on AdS space, where numerical methods can be used to construct magnetic bags with a large range of values of $n$ \cite{Sutcliffe:2011sr}.  Thus it is widely believed that magnetic bags exist for infinitely many values of $n$, and that they are the most tightly-packed configurations of monopoles.

The magnetic charge $q=4\pi n/e$ of a magnetic bag remains finite in the limit $n\to\infty$ provided one takes a double-scaling limit $e\to\infty$ such that $n/e$ remains finite.  In this limit the BPS energy $E=vq$ and also the size of the bag remain finite.  The double scaling limit causes two of the three $\asu(2)$ components of the fields to be exponentially suppressed. This can be seen from the D-brane interpretation discussed in section \ref{dbraneint}, where the `W-boson' strings stretching between different D-branes have masses which diverge as $\sim e||\Phi||$. With just one generator of $\asu(2)$ left, we have $\au(1)$ valued fields, which we denote $\phi$ and $f$. Explicitly, we have 
\begin{equation}
\Phi\rightarrow~\di\begin{pmatrix}\phi&&0\\0&&-\phi\end{pmatrix}\eand F\rightarrow~\di\begin{pmatrix}f&&0\\0&&-f\end{pmatrix}
\end{equation}
in local gauges as $n\to\infty$.  The surface of a bag becomes infinitely thin as $n,e\to\infty$, and can be represented by a surface $S\subset\FR^3$.  One has $f=0$ inside the bag, and hence that $\phi$ is constant; $\phi$ is continuous on $S$, but $f$ is not.  We will assume that $\phi=0$ inside the bag.

We will be concerned with magnetic bags only in this abelian double-scaling limit.  One could take similar limits of walls or chains: here one sends the topological charge per unit area (or length) and the coupling constant $e$ to infinity, in such a way that the magnetic charge per unit area (or length) stays finite.  The limiting configuration for walls could have a discontinuity in $f$ along a plane, while for chains one could perhaps arrange for a singularity along a line or a discontinuity on a cylinder.

All of these abelian limiting configurations are examples of what we will refer to as {\em magnetic domains} $\Omega$ in three dimensions: These are monopole configurations characterized by continuous $\au(1)$-valued fields $(f,\phi)$ satisfying the following properties:
\begin{itemize}
 \setlength{\itemsep}{-1mm}
 \item $f$ is closed, and therefore we have locally a gauge potential $a$ with $f=\dd a$,
 \item $f$ and $\phi$ satisfy the Bogomolny monopole equation $f=*\dd \phi$ in the region $\Omega\subset \FR^3$,
 \item $\dd \phi\neq 0$ in $\Omega$ and
 \item depending on the shape and dimensionality of the boundary of the domain $\Omega$, $\phi$ satisfies certain boundary conditions.
\end{itemize}

\subsection{Nahm transform and the fuzzy funnel}\label{sect:FuzzyFunnel}

The correspondence between monopoles and Nahm data is well known; Stated formally, we have
\begin{thm}
\label{thm}
\cite{Nahm:1981nb,Hitchin:1983ay} Up to gauge equivalence, there is a one-to-one correspondence between\vspace*{-0.2cm}
\begin{itemize}
 \setlength{\itemsep}{-1mm}
 \item solutions to the Bogomolny equation $F=*\dd_A\Phi$ on $\FR^3$ with the boundary conditions
\begin{eqnarray}
||\Phi||\rightarrow v~,~~~\frac{\pa ||\Phi||}{\pa\Omega}=\CO(r^{-2})~,~~~||\dd_A\Phi||=\CO(r^{-2})
\end{eqnarray}
as $r\rightarrow \infty$ and
 \item solutions to the Nahm equations
\begin{equation}\label{eq:NahmFinite}
\frac{\dd T^i}{\dd s} = \frac{e}{2}\,\eps_{ijk}[T^j,T^k]~,
\end{equation}
satisfying the reality and boundary conditions
\begin{equation}
\label{RC}
T^i(-s)=T^i(s)^t~,~~
T^i(s) = \frac{1}{e} \frac{J^i}{v-s} + \CO(1)~~\mbox{as}~~s\rightarrow v~,
\end{equation}
where $T^i\in\au(n)\times\CC^\infty(-v,v)$ and $J^i\in\au(n)$ form an $n$-dimensional irreducible representation of $\asu(2)$.
\end{itemize}
\end{thm}

Instead of regarding the $T^i$ as functions on the interval $\CI=(-v,v)$ taking values in $\au(n)$, we can interpret them as functions on $S^2_F\times \CI$, where $S^2_F$ is a fuzzy sphere at level $n$. To understand this statement, let us briefly recall the Berezin-Toeplitz quantization of the 2-sphere \cite{Berezin:1974du}, see also \cite{Balachandran:2005ew,IuliuLazaroiu:2008pk} and references therein. We start from the round sphere $S^2\cong\CPP^1$ endowed with its Fubini-Study metric and the corresponding K\"ahler form $\omega$. As usual in geometric quantization, we have to pick an ample line bundle (the prequantum line bundle), from whose global sections we derive a Hilbert space $\CCH_n$. We choose the line bundle $L_n=\CO(n-1)$ with first Chern number $c_1=n-1$ and we will moreover work with K\"ahler polarization. This means that the Hilbert space is given by the global holomorphic sections of $L_n$: \begin{equation}
\CCH_n=H^0(\CPP^1,L_n)\cong \FC^n~.
\end{equation}
Using the volume form $\omega$, one can construct an inner product on $\CCH_n$ via
\begin{equation}
 \langle s_1|s_2\rangle:=\int_{\CPP^1} \frac{\omega(z,\zb)}{(1+z\zb)^n}~ \overline{s_1(z)}s_2(z)~,
\end{equation}
where $z\in \FC\cup\{\infty\}$ denotes a point on $\CPP^1$. Moreover, we can construct an overcomplete set of coherent states $|z\rangle\in \CCH_n$ for each $z$. These are used in the definition of the coherent state projector $P_{z,\zb}:=\frac{|z\rangle\langle z|}{\langle z| z \rangle}$, which provides a bridge between the classical and the quantum world, as $P_{z,\zb}\in\CC^\infty(\CPP^1)\otimes\sEnd(\CCH_n)$. We define the {\em Berezin symbol map}
\begin{equation}
 \sigma_n: \sEnd(\CCH_n)\rightarrow \CC^\infty_n(\CPP^1)\subset \CC^\infty(\CPP^1)\ewith \sigma_n(A):=\tr(P_{z,\zb} A)~,
\end{equation}
and the {\em Toeplitz quantization map}
\begin{equation}
 \CCT_n: \CC^\infty(\CPP^1)\rightarrow \sEnd(\CCH_n)\ewith \CCT_n(f):=\int_{\CPP^1}\omega(z,\zb) f(z,\zb)P_{z,\zb}~.
\end{equation}
The set $\CC^\infty_n(\CPP^1)$ is called the {\em set of quantizable functions at level $n$}. Both the above maps combine to the Berezin transform $\beta_n:\CC^\infty(\CPP^1)\rightarrow \CC^\infty_n(\CPP^1)$, where $\beta_n(f)=\sigma_n(\CCT_n(f))$. We now have the following results in the large $n$ limit \cite{Bordemann:1993zv}, see also \cite{Schlichenmaier:1998mv}:
\begin{equation}\label{eq:approximations}
 \|\di n[\CCT_n(f),\CCT_n(g)]-\CCT_n(\{f,g\})\|=\CO\left(\frac{1}{n}\right)\eand
 \beta_n(f)(z,\zb)=f(z,\zb)+\CO\left(\frac{1}{n}\right)~.
\end{equation}

On the set of quantizable functions $\CC^\infty_n(\CPP^1)$, we can invert $\sigma_n$ to obtain a quantization map $\sigma^{-1}_n$ from real functions in $\CC^\infty_n(\CPP^1)$ to $\au(n)$, the set of real endomorphisms on $\CCH_n$. In this quantization procedure, $n$ plays essentialy the role of $1/\hbar$. The fuzzy sphere is now defined via its algebra of functions $\sEnd(\CCH_n)\cong \CC^\infty_n(\CPP^1)$. Note that the operator product on $\sEnd(\CCH_n)$ induces a ``star product'' on $\CC^\infty_n(\CPP^1)$ by $f\star g=\sigma_n^{-1}(\sigma_n(f)\sigma_n(g))$.

Explicitly, the coordinate functions $x^i$ describing the embedding $S^2\subset \FR^3$ are mapped to the operators $X^i:=\frac{2\di J^i}{n}\in \au(n)$, where $J^i$ form an $n$-dimensional irreducible representation of $\asu(2)$. For these, we have the identities
\begin{eqnarray}
\label{eq:4.1}
X^iX^j-X^jX^i &=& \frac{2\di}{n}\eps_{ijk}X^k~,\\
\label{eq:4.2}
(X^1)^2+(X^2)^2+(X^3)^2 &=& 1 - \frac{1}{n^2}~,
\end{eqnarray}
which makes the limit $S^2_F\rightarrow S^2\subset \FR^3$ as $n\rightarrow\infty$ clear. 

General functions in $\CC^\infty_n(S^2)$ split up into representations of the rotation group $\sSO(3)\simeq \sSU(2)$. These representations are given by the spherical harmonics $Y_{\ell m}$, labeled by integers $0\leq \ell< n$, $m\in \RZ$ with $|m|\leq \ell$:
\begin{equation}
\CC_n^\infty(S^2)= \bigoplus_{\ell=0}^{n-1} \bigoplus_{m=-\ell}^{\ell}Y_{\ell m}\cong \bigoplus_{i=1}^n ({\bf 2i-1})_\FR~.
\end{equation}
Here, ${\bf i}$ is the $i$-dimensional irreducible representation of $\asu(2)$. A subscript $\FR$ denotes projection onto its real part under the obvious antilinear involution. Note that the functions $Y_{1m}$, $m=-1,0,1$ are linear combinations of the coordinate functions $x^1,x^2,x^3$. Under quantization, elements of $\CC_n^\infty(S^2)$ are mapped to general elements of $\au(n)$, which form the same sums of representations of $\asu(2)\cong \aso(3)$:
\begin{equation}
(\sEnd(\FC^n))_\FR\cong\au(n)\cong(\overline{{\bf n}}\otimes{\bf n})_\FR= \bigoplus_{i=1}^n ({\bf 2i-1})_\FR~,
\end{equation}

In the limit $n\rightarrow\infty$, the fuzzy sphere $S_F^2$ becomes the ordinary sphere $S^2$ and $\overline{\CC^\infty_n(S^2)}\rightarrow \CC^\infty(S^2)$. As implied by \eqref{eq:approximations}, the Lie bracket on $\au(n)$ goes over to the Poisson bracket on $\CC^\infty(S^2)$. All this suggests that in the case of magnetic bags, for which $n\rightarrow \infty$, the Nahm data should be extended from functions on an interval $\CI$ taking values in $\au(n)$ to functions on $S^2\times\CI$. A Nahm construction using this point of view has been developed in \cite{Harland:2011tm}. In the following, we will also allow for other 2-manifolds such as $\FR^2$ and $\FR\times S^1$ to replace $S^2$ and thus extend this construction to a large class of magnetic domains.

\subsection{Nahm transform for magnetic domains}\label{sect:NahmTrans}

We start from a real two-dimensional manifold $M$ without boundary. A volume form $\omega$ on $M$ induces a symplectic structure, which in turn leads to a Poisson bracket on $\CC^\infty(M)$. This Poisson bracket can be trivially extended to a Poisson bracket $\{\cdot,\cdot\}_\omega$ on $\CC^\infty(M\times \CI)$, where $\CI$ is the union of finitely many intervals on the positive real line. We denote the resulting Poisson algebra by $\Pi_\omega$. 

By {\em $\Pi_\omega$-valued Nahm data} or {\em $\Pi_\omega$-Nahm data} for short, we understand a triple of functions $t^i\in \CC^\infty(M\times \CI)$, which satisfy the $\Pi_\omega$-Nahm equation
\begin{equation}\label{eq:NahmInfinity}
 \frac{\partial t^i}{\partial s} = \frac{4\pi}{q}\frac{1}{2}\eps_{ijk} \{t^j,t^k\}_\omega~.
\end{equation}

Below we will state and prove a theorem which shows how solutions of the $\Pi_\omega$-Nahm equation can be used to construct magnetic domains.  However, before doing so we need to introduce the concept of the volume type of a volume form on a 2-manifold.  In general a non-compact manifold $M$ may be written as a union of a compact subset $K$ and a collection of open sets $U$, called ends.  For example $\FR^2\backslash\{(0,0)\}$ has two ends, one near $r=\infty$ and one near $r=0$:
\begin{equation}
\begin{aligned}
\FR^2\backslash\{(0,0)\} &= U_0\cup K\cup U_\infty~,\\
U_0 = \{x^ix^i < 1 \}~,~~~K &= \{x^ix^i = 1 \}~,~~~U_\infty = \{x^ix^i > 1 \}~.
\end{aligned}
\end{equation}
Given any volume form $\omega$ on $M$ one may measure the volume of each of its ends, and this could be either infinite or finite.  We say that two volume forms $\omega_1$, $\omega_2$ have the same {\em volume type} if every end $U$ has either infinite $\omega_1$-volume and infinite $\omega_2$-volume, or finite $\omega_1$-volume and finite $\omega_2$-volume.  The notion of volume type is independent of the choice of compact set $K$ provided that $K$ is big enough -- we refer the reader to the appendix or reference \cite{Greene:1979aa} for more details. Note that two volume forms on a compact manifold are trivially of the same volume type.

As a simple example, consider the following two volume forms on $\FR^2\backslash\{(0,0)\}$:
\begin{equation}
\omega_1 = \dd x^1 \wedge \dd x^2~,~~~\omega_2 = \frac{\dd x^1 \wedge \dd x^2}{x^ix^i}~.
\end{equation}
These do not have the same volume type, since the volume of $U_0$ is finite when measured with $\omega_1$ but infinite when measured with $\omega_2$.  Note however that both volume forms give $M$ infinite volume.

It is not hard to show that two volume forms related by a diffeomorphism that does not permute the ends have the same volume type.  The converse statement was proven in \cite{Greene:1979aa}: if $\omega_1,\omega_2$ have the same volume type then there exists a diffeomorphism $u$ of $M$ such that $u^*\omega_2=\omega_1$.

\begin{thm}
\label{thm:2}
Up to gauge equivalence, there is a one-to-one correspondence between
\vspace{-0.2cm}
\begin{itemize}
 \setlength{\itemsep}{-1mm}
 \item sets of $\Pi_\omega$-Nahm data with the property that the map from $M\times \CI$ to $\Omega\subset\FR^3$ defined by the $t^i$ is a diffeomorphism $t:M\times \CI\rightarrow \Omega$, and
 \item magnetic domains $\Omega$ that are diffeomorphic to $M\times \CI$, where the restriction of $f$ to any slice $M\times\{s_0\}$ has the same volume type as $\omega$ and $\CI$ is the range of $\phi$. Explicitly, there is a diffeomorphism $u:\Omega\rightarrow M\times \CI$ such that $f=\frac{q}{4\pi}u^*\omega$ and $\phi=s\circ u$ on $\Omega$. 
\end{itemize}
\end{thm}

The proof follows closely that given in \cite{Harland:2011tm} for magnetic bags, and is similar to one given in \cite{Dunajski:2003ep}: The $\Pi_\omega$-Nahm data provide us with a diffeomorphism $t$ and its inverse $u$,
\begin{equation}\label{eq:diff}
M\times \CI ~\overset{t}{\underset{u}\rightleftarrows}~\Omega\subset\FR^3~.
\end{equation}
We will use local coordinates $\theta^{1,2}$ on $M$, $s$ on $\CI$ and Cartesian coordinates $y^i$ on $\Omega\subset \FR^3$. By definition of the Poisson bracket, the Nahm equation \eqref{eq:NahmInfinity} is equivalent to 
\begin{equation}
\dd t^i\wedge\omega=\frac{4\pi}{q}\frac{1}{2}~\eps_{ijk}~\dd t^j \wedge \dd t^k\wedge \dd s ~,
\end{equation}
where $\omega$ is the volume form on $M$. This is an equation on $M\times \CI$, which we want to pull back along $u$ to an equation on $\Omega\subset\FR^3$, identifying $y^i=u^* t^i$:
\begin{equation}
\begin{aligned}
\dd y^i\wedge u^*\omega=\frac{4\pi}{q}\frac{1}{2}\eps_{ijk}\dd y^j \wedge \dd y^k\wedge u^*\dd s&=\frac{4\pi}{q}* \dd y^i\wedge u^*\dd s=\frac{4\pi}{q} \dd y^i\wedge *~u^*\dd s\\
\iff~\frac{q}{4\pi}u^*\omega&=*\dd~ u^*s~,
\end{aligned}
\end{equation}
and therefore $f=*\dd \phi$. Note that $\omega$ is a volume form on $M$ and therefore closed. This means that locally, there exists a gauge potential $a$ such that $f=\dd a$.

Alternatively, we can start from the fields $f:=\frac{q}{4\pi}u^*\omega$ and $\phi:=u^*s$ and determine the conditions necessary for the Bogomolny equation $f=*\dd \phi$ to hold. For this, we pull back $f=*\dd \phi$ to $M\times \CI$ to get 
\begin{equation}
\frac{q}{4\pi}~\omega= t^**\dd\phi~. 
\end{equation}
We compute 
\begin{equation}
 t^**\dd\phi=\eps^{ijk}  \frac{1}{2}\frac{\partial s}{\partial t^i}\left(\frac{\partial t^j}{\partial \theta^a}\frac{\partial t^k}{\partial  s}\dd\theta^a\wedge\dd s+\frac{\partial t^j}{\partial \theta^a}\frac{\partial t^k}{\partial  \theta^b}\dd\theta^a\wedge\dd\theta^b\right)~.
\end{equation}
When the $\Pi_\omega$-Nahm equation holds, the unwanted term $\eps^{ijk}\frac{\partial s}{\partial t^i}\frac{\partial t^j}{\partial \theta^a}\frac{\partial t^k}{\partial  s}$ vanishes since\linebreak $\frac{\partial s}{\partial t^i}\frac{\partial t^i}{\partial \theta^a}=\frac{\partial s}{\partial \theta^a}=0$ and the remaining term gives $\frac{q}{4\pi}\omega$. With a little more work, it can be shown that the Nahm equation is in fact equivalent to $f=*\dd \phi$. We will use a similar argument when discussing the loop space self-dual string bags in section \ref{sect:LSSDS}.

The inverse construction is done for each connected component in $\Omega$ separately. Let us therefore restrict to one connected component $\Omega^c$ of $\Omega$, on which the range of $\phi$ is given by some interval $\CI$. By assumption, the magnetic domain $\Omega^c$ is diffeomorphic to $M\times\CI$ with $\dd\phi\neq0$ everywhere. The direct product structure $M\times\CI$ translates into a foliation of $\Omega^c$ by two-dimensional surfaces $\Sigma_\phi$ that are diffeomorphic to $M$. These surfaces are formed by the level sets of $\phi$. We pick an element $\phi_0=s_0\in \CI$ and the corresponding level set $\Sigma_{\phi_0}=\{p\in \Omega^c|\phi(p)=\phi_0\}$ together with the embedding $i: \Sigma_{\phi_0}\embd \Omega^c$. Now $f$ and $\omega$ have the same volume type, so the non-compact version of Moser's theorem \cite{Greene:1979aa} implies that there is a diffeomorphism $w:M\rightarrow \Sigma_{\phi_0}$ and a constant $q\in\FR$ such that $w^*i^*f=\frac{q}{4\pi}\omega$ (see also \cite{
0198504519}).  We will now extend the map $i$ to a diffeomorphism $t:\CI\times M\rightarrow \Omega^c$ as done in \cite{
Harland:2011tm}: The vector field $\der{s}$ has the properties
\begin{equation}
 \CL_{\der{s}}s=1\eand \iota_{\der{s}}\omega=0~,
\end{equation}
where $\CL$ denotes the Lie derivative. On $\Omega^c$, we have analogously the normalized gradient of $\phi$, i.e.\ the vector field
\begin{equation}
 Y=\left(\derr{\phi}{y^j}\derr{\phi}{y_j}\right)^{-1}\derr{\phi}{y^i}\der{y_i}~,
\end{equation}
which satisfies
\begin{equation}
 \CL_Y\phi=1\eand \iota_Y f=0~.
\end{equation}
We now solve the differential equations
\begin{equation}
 \dderr{y^i}{s}=Y^i(y(s))
\end{equation}
with the boundary condition $y(s_0)=w\circ i$ at $s_0=\phi_0$. The map $y$ yields a diffeomorphism between $\tilde{\CI}\times M$ and $\Omega^c$, where $\tilde{\CI}$ is some interval in $\FR$ containing $s_0$. Because of
\begin{equation}
 \dderr{\phi(y^i(s))}{s}=\dderr{y^i}{s}\derr{\phi}{y^i}=\CL_Y\phi=1~,
\end{equation}
$\tilde{\CI}$ is identical to the range of $\phi$ and therefore to $\CI$, and we can identify $t$ with $y$. The one-to-one correspondence is then shown by composing the transform with the inverse transform to get the identity. This completes the proof.

\

The fact that we can find a prescription for the explicit construction of magnetic domains reflects that they are described by integrable equations. They therefore come with an infinite number of conserved charges as shown in \cite{Harland:2011tm} for magnetic bags. 

The boundary conditions imposed on the $\Pi_\omega$-Nahm data at the edges of the intervals contained in $\CI$ are in direct correspondence to the boundary conditions of the fields describing the magnetic domain, as we will show in detail for various examples in the next section.

\subsection{Examples}\label{sec:3DExamples}

For magnetic bags, the boundary $S$ of the domain $\Omega$ is diffeomorphic to a sphere. The domain $\Omega$ itself is then 
$\FR^3$ with the interior of $S$ excluded. The boundary conditions imposed are that $\phi=0$ on $S$ and $\phi$ tends to some positive constant $v$ as $r\to\infty$.  Due to the Bogomolny equation \eqref{eq:BogomolnyMonopole}, the Higgs field $\phi$ is a harmonic function on $\Omega$ and furthermore, because of the definition of $q$ in \eqref{magnetic charge}, $\phi$ has the following asymptotic expansion:
\begin{equation}\label{eq:HiggsB}
 \phi\sim v - \frac{q}{4\pi r}+\CO(1)~~~\mbox{for}~~~r\rightarrow \infty~.
\end{equation}
The $\Pi_\omega$-Nahm data are now functions of $S^2\times\CI$, where $\CI=[0,v)$. The lower bound of $\CI$ corresponds to the surface $S$, while the upper bound of $\CI$ corresponds to $S^2_\infty$, the boundary of $\FR^3$ at infinity. This asymptotic behavior of the Higgs field \eqref{eq:HiggsB} induces the following boundary condition on the $\Pi_\omega$-Nahm data:
\begin{equation}
t^i(x,s) = \frac{q}{4\pi} \frac{ x^i}{v-s}+\CO(1)~~~\mbox{as}~~~s\rightarrow v~.
\end{equation}

The simplest example for a magnetic bag is the spherical one. It has $\Pi_\omega$-Nahm data \cite{Harland:2011tm}
\begin{equation}\label{eq:NDsphbag}
t^i(x,s) = \frac{q}{4\pi} \frac{ x^i}{v-s} ~~,
\end{equation}
where $ x \in S^2\subset \FR^3$. The inverse map $u:\Omega\rightarrow S^2\times I$ is 
\begin{equation}
u(\vec y)=\left(\frac{\vec y}{r},v-\frac{q}{4\pi r}\right)~,
\end{equation}
from which we compute
\begin{equation}
\phi=u^* s =\begin{cases} v-\frac{q}{4\pi r} & r\geq \frac{q}{4\pi v} \\ 0 & r < \frac{q}{4\pi v} \end{cases} ~~. 
\end{equation}
Thus, $\Omega$ is given by $\{\vec{y}~|~|\vec{y}|\geq \frac{q}{4\pi v}\}\subset \FR^3$. Now on $S^2$ we have $\omega=\sin\theta^1\dd\theta^1\wedge\dd\theta^2=\frac{1}{4}\eps_{ijk} x^i \dd  x^j\wedge\dd  x^k$ and so
\begin{equation}
\begin{aligned}
 f=&\frac{q}{4\pi}u^*\omega=\begin{cases} \frac{q}{8\pi r^3}\eps_{ijk}y^i\dd y^j\wedge\dd y^k & r\geq \frac{q}{4\pi v} \\ 0 & r < \frac{q}{4\pi v} \end{cases}~~.
\end{aligned}
\end{equation}
These fields satisfy $f=*\dd\phi$ on $\Omega$.

A generalization of this example is the ellipsoidal bag, stretched in the $y^3$ direction, for which the $\Pi_\omega$-Nahm data reads as 
\begin{equation}
 t(x,s)=\frac{q\lambda}{4\pi} \left(\frac{ x^{1}}{\sinh(\lambda(v-s))},\frac{ x^{2}}{\sinh(\lambda(v-s))},\frac{ x^3}{\tanh(\lambda(v-s))} \right)~.
\end{equation}
In the limit $\lambda\rightarrow 0$, the $\Pi_\omega$-Nahm data reduce to the spherical case \eqref{eq:NDsphbag}. Let us now restrict to $\lambda=1$ for simplicity. 

The inverse map $u:\Omega\rightarrow S^2\times \CI$ is 
\begin{equation}
u(\vec y)=\left(\left(\frac{y^1}{\alpha},\frac{y^2}{\alpha},y^3\sqrt\frac{ p}{p\alpha^2+1}\right),v-\sinh^{-1}\left(\frac{1}{\sqrt{p}\alpha}\right)\right)~,
\end{equation}
where $p=(4\pi/q)^2$ and 
\begin{equation}
\alpha^2=\frac{p r^2-1+\sqrt{4p((y^1)^2+(y^2)^2)+(pr^2-1)^2}}{2p}~.
\end{equation}
Therefore $\phi(\vec y)=v-\sinh^{-1}(\frac{q}{4\pi\alpha(\vec y)})$ and $f=\eps_{ijk} \frac{1}{\alpha\sqrt{p\alpha^2+1}}\frac{\pa\alpha}{\pa y^i}\dd y^j\wedge\dd y^k$ on
\begin{equation}
 \Omega := \left\{ \vec{y}~\Big|~(y^1)^2+(y^2)^2+\frac{1}{\cosh^2 v} (y^3)^2\geq \frac{1}{p\sinh^2 v}\right\}\subset \FR^3~.
\end{equation}

We can also consider a circular disc, i.e.\ a degenerate magnetic bag completely squashed in the $y^3$ direction, with $\Pi_\omega$-Nahm data 
\begin{equation}
 t(x,s)=\frac{q}{8v} \left(\frac{ x^{1}}{\sin(\pi(v-s)/2v)},\frac{ x^{2}}{\sin(\pi(v-s)/2v)},\frac{ x^3}{\tan(\pi(v-s)/2v)} \right)
\end{equation}
and inverse 
\begin{equation}
u(\vec y)=\left(\left(\frac{y^1}{\alpha},\frac{y^2}{\alpha},y^3\sqrt\frac{ p}{p\alpha^2-1}\right),v-\frac{2v}{\pi}\sin^{-1}\left(\frac{1}{\sqrt{p}\alpha}\right)\right)~,
\end{equation}
where $p=(8v/q)^2$ and 
\begin{equation}
\alpha^2=\frac{p r^2+1+\sqrt{-4p((y^1)^2+(y^2)^2)+(pr^2+1)^2}}{2p}~.
\end{equation}
Therefore $\phi=v(1-\frac{2}{\pi}\sin^{-1}(\frac{q}{8v\alpha(\vec y)}))$ and $f=\eps_{ijk} \frac{1}{\alpha\sqrt{p\alpha^2-1}}\frac{\pa\alpha}{\pa y^k}\dd y^j\wedge\dd y^k$ on $\Omega=\FR^3\backslash D$, where $D$ is a disc in the $y^1$-$y^2$-plane with radius $q/8v$. The Higgs field $\phi$ (and $-\phi$) are used in the plots in Figure 1. These plots will find a natural interpretation in terms of D3-branes as explained in section \ref{sec:branes}.
\begin{figure}[h]
\center
\begin{picture}(420,100)
~~~~\includegraphics[width=50mm]
{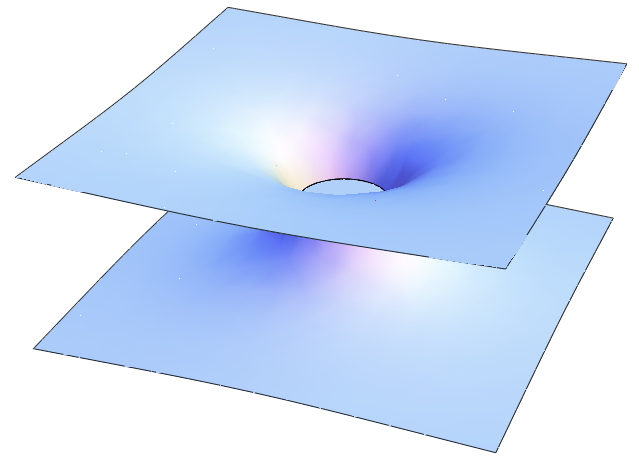}~~~~~~~~\includegraphics[width=50mm]{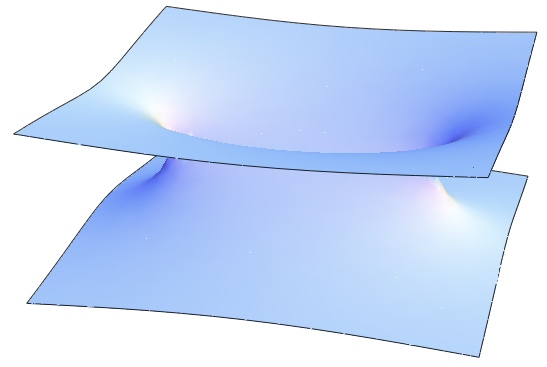}
\put(30.0,35.0){\vector(1,0){20}}
\put(30.0,35.0){\vector(0,1){20}}
\put(30.0,35.0){\vector(-1,-3){6}}
\put(70.0,35.0){\makebox(0,0)[c]{$x^1,x^2$}}
\put(30.0,62.0){\makebox(0,0)[c]{$s$}}
\put(30.0,14.0){\makebox(0,0)[c]{$x^3$}}
\end{picture}
\caption{Plots of the Higgs field $\phi$ (and $-\phi$) for the spherical magnetic bag and the circular magnetic disc. The vertical axis is the s-direction and one of the circular symmetric directions is suppressed.}
\end{figure}
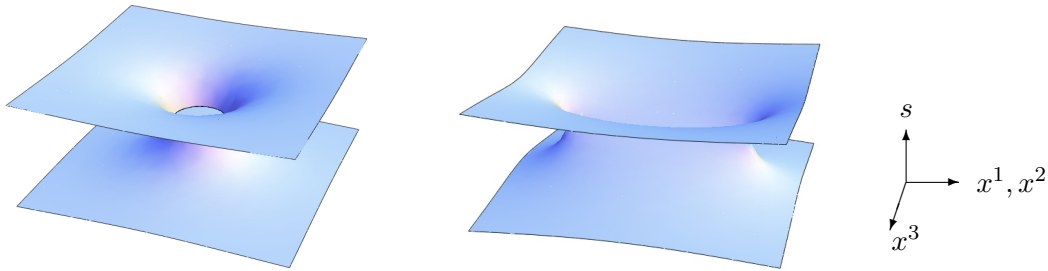

The flat magnetic wall \cite{Lee:1998isa} arises from a map $t:\FR^2\times \CI\rightarrow \FR^3_{y^3> 0}$, where $\CI=[0,\infty)$. The Poisson bracket on $\FR^2$, arising from the symplectic form $\omega=\dd x^1\wedge\dd x^2$, is just 
\begin{equation}
\{x^a,x^b\}=\eps^{ab}~,~~~a,b=1,2~.
\end{equation}
The $\Pi_\omega$-Nahm data for the flat wall \cite{Harland:2011tm} are
\begin{equation}
t^a(x,s)=x^a~,~~t^3(x,s)= \frac{4\pi}{q} s~,
\end{equation}
and the inverse map is
\begin{equation}
u(\vec y)=\left( (y^1,y^2),\frac{q}{4\pi} y^3\right)~.
\end{equation}
This gives solutions to the Bogomolny equation 
\begin{equation}
\phi=u^* s= \frac{q}{4\pi} y^3 ~,~~f=\frac{q}{4\pi} u^* \omega =\frac{q}{4\pi} \dd y^1\wedge \dd y^2~.
\end{equation}

Note that the Higgs field is a harmonic function on $\Omega$, which is independent of $y^1$ and $y^2$ as expected. More general magnetic walls, correspondingly, would still have $\Pi_\omega$-Nahm data $t:\FR^2\times \CI\rightarrow \FR^3_{y^3> 0}$ satisfying the boundary condition
\begin{equation}
t^3(x,s)\sim \frac{4\pi}{q} s ~~\mbox{as }s\rightarrow\infty~.
\end{equation}

Finally, we can also consider a magnetic tube along the $y^3$ axis. This arises from a map $t:S^1\times\FR\times \CI\rightarrow \Omega\subset\FR^3$. The Poisson bracket on $S^1\times\FR$, induced by the symplectic form $\omega=\eps^{ab}x^a\dd x^b\wedge\dd z$, is
\begin{equation}
\{x^1,x^2\}=0~,~~\{z,x^a\}=\eps^{ab}x^b~,
\end{equation}
where $(x^1,x^2)\in S^1\subset\FR^2$. The $\Pi_\omega$-Nahm data are given by
\begin{equation}
t^a(x,z,s)=\de^{\frac{4\pi}{q}(s-v)}x^a~,~~t^3(x,z,s)=z~,~~~a=1,2~,
\end{equation}
and the inverse map is
\begin{equation}
u(\vec y)=\left( \left(\frac{y^1}{r},\frac{y^2}{r}\right),y^3,\frac{q}{4\pi}\ln(r)+v\right)~,
\end{equation}
where $r^2:=(y^1)^2+(y^2)^2$. From here we can see that the bag surface is a cylinder along the $y^3$-axis with radius $r=\de^{-\frac{4\pi}{q}v}$ and $\Omega$ is the exterior of this cylinder in $\FR^3$.

This gives solutions to the Bogomolny equation 
\begin{equation}
\phi=u^* s= \frac{q}{4\pi}\ln(r)+v ~,~~f=\frac{q}{4\pi} u^* \omega =\eps^{ab}\frac{q}{4\pi}\frac{y^a}{r^2}\dd y^b\wedge \dd y^3~.
\end{equation}

General magnetic tubes would have $\Pi_\omega$-Nahm data with the boundary condition 
\begin{equation}
t^a(x,z,s)\sim\de^{\frac{4\pi}{q}(s-v)}x^a~~~\mbox{for }~a=1,2~~\mbox{as }s\rightarrow\infty~.
\end{equation}
\begin{figure}[h]
\center
\begin{picture}(420,70)
\includegraphics[width=45mm]{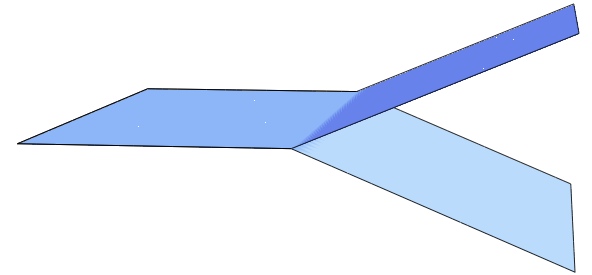}~~~~~~~~~~~~~~~~~~~~\includegraphics[width=45mm]{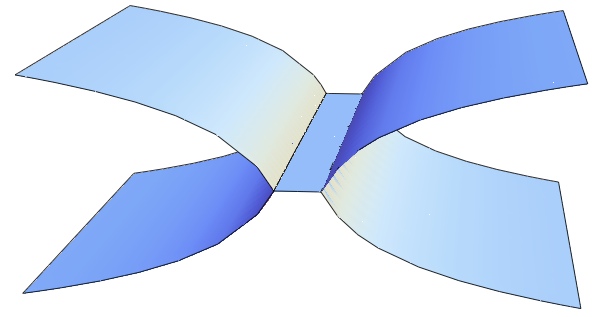}
\put(30.0,35.0){\vector(1,0){20}}
\put(30.0,35.0){\vector(0,1){20}}
\put(30.0,35.0){\vector(-1,-3){6}}
\put(70.0,35.0){\makebox(0,0)[c]{$x^1,x^2$}}
\put(30.0,62.0){\makebox(0,0)[c]{$s$}}
\put(30.0,14.0){\makebox(0,0)[c]{$x^3$}}
\put(-180.0,35.0){\vector(1,0){20}}
\put(-180.0,35.0){\vector(0,1){20}}
\put(-180.0,35.0){\vector(-1,-3){6}}
\put(-150.0,35.0){\makebox(0,0)[c]{$x^3$}}
\put(-180.0,62.0){\makebox(0,0)[c]{$s$}}
\put(-180.0,10.0){\makebox(0,0)[c]{$x^1,x^2$}}
\end{picture}
\caption{Magnetic wall and magnetic tube. The vertical axis is the s-direction and one of the symmetric directions of $\FR^3$ is suppressed.}
\end{figure}
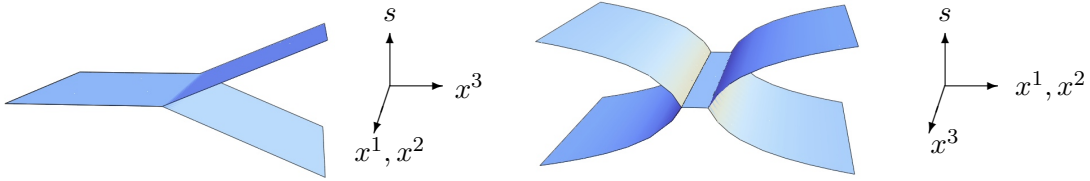

\subsection{Magnetic domains as limits of monopole configurations}\label{commentsConjecture}

As stated above, abelian magnetic bags are expected to correspond to the large $n$ limits of non-abelian magnetic monopoles.  More precisely, Bolognesi has made the following conjecture \cite{Bolognesi:2005rk}, cf.\ \cite{Harland:2011tm}:

\begin{conj}\label{conj:Bolognesi}
 For any magnetic bag $(f,\phi)$, there is a sequence $(A^{(n)},\Phi^{(n)})$ of charge $n$ solutions to the Bogomolny monopole equations $F^{(n)}=\dd_{A^{(n)}} A^{(n)}=e_n\,\star \dd \Phi^{(n)}$ with coupling constant $e_n\in\FR$ and gauge group $\sSU(2)$, such that in the limit $n\rightarrow \infty$:
 \begin{equation}\label{eq:limits}
  2\pi \, \frac{n}{e_n}\rightarrow q~,~~~\|\Phi^{(n)}\|\rightarrow\phi\eand-\frac{\tr(F^{(n)}\Phi^{(n)})}{2\|\Phi^{(n)}\|}\rightarrow f~.
 \end{equation}
\end{conj}

Recall that the ADHMN construction gives a one-to-one correspondence between gauge equivalence classes of sets of Nahm data and gauge equivalence classes of solutions to the Bogomolny monopole equations. Note that the limits \eqref{eq:limits} in conjecture \ref{conj:Bolognesi} are gauge invariant. This suggests that if conjecture \ref{conj:Bolognesi} is true, then for each set of $\Pi_\omega$-Nahm data $(t^i(s))$ corresponding to a magnetic bag $(f,\phi)$, one can find a sequence of Nahm data $(T^i_{(n)}(s))$ for finite-charge monopoles that converges towards $t^i$ in the large $n$ limit. Moreover, the solutions $(T^i_{(n)}(s))$ can be extended to the full interval $\CI_2=\CI\cup -\CI$. Let us be more precise:

\begin{conj}\label{conj:reducedConj}
 For each solution $(t^i)$, $t^i\in\CC^\infty(S^2\times \CI)$ of the infinite-charge Nahm equation \eqref{eq:NahmInfinity} corresponding to a magnetic bag, there is a sequence of solutions $(T^i_{(n)})$, $T^i_{(n)}\in \au(n)\otimes \CC^\infty(\CI_2)$ of the finite-charge Nahm equation \eqref{eq:NahmFinite} such that in the limit $n\rightarrow \infty$: $\sigma_n(T^i_{(n)}(s))\rightarrow t^i(s)$ on $\CI$. Here, $\sigma_n$ is the Berezin symbol map $\sigma_n:\au(n)\rightarrow \CC_n^\infty(S^2)$ introduced above.
\end{conj}

To find a sequence of sets of Nahm data $T^i_{(n)}(s)$ converging towards a set of $\Pi_\omega$-Nahm data $t^i(s)$ for a magnetic bag, one would ideally like a non-trivial Lie algebra homomorphism from the Poisson algebra $\CC^\infty(S^2)$ to the Lie algebra $\au(n)$. However, such a map does not exist. The best one can do is to use an approximate Lie algebra homomorphisms, just as the Toeplitz quantization map, cf.\ \eqref{eq:approximations}. 

First, it is necessary to extend the $\Pi_\omega$-Nahm data for magnetic bags from the half-interval $\CI=[0,v)$ to the full interval $\CI_2=(-v,v)$.  The operation of transposition on a matrix can be interpreted as the operation of a reflection $R\in \mathrm{O}(3)$ on the fuzzy sphere \cite{Harland:2011tm}, so the reality condition $T^i(-s)=T^i(s)^t$ for monopole Nahm data should be replaced by the condition $t^i(x,-s)=t^i(Rx,s)$ for bag $\Pi_\omega$-Nahm data.  Thus $\Pi_\omega$-Nahm data on $\CI$ can be extended to $\CI_2$, but doing so may introduce a discontinuity at $s=0$.

The discontinuity is not present if the $\Pi_\omega$-Nahm data satisfy $t^i(Rx,0)=t^i(x,0)$.  If this is the case the corresponding magnetic bag will be degenerate, in the sense that the volume contained inside the magnetic bag will vanish.  It is not hard to convince oneself that Bolognesi's conjecture is true for these degenerate bags, at least in the form of conjecture \ref{conj:reducedConj}: to obtain Nahm data for a monopole corresponding to a degenerate bag, one only needs to take $T^i_{(n)}(0):=\CCT_n(t^i(0))$ as initial conditions and solve the Nahm equation.  The Nahm equation implies that the condition $T^i(-s)_{(n)}=T^i(s)_{(n)}^t$ is automatically satisfied, because the matrices $T^i_{(n)}(0)$ are by construction symmetric. As the failure of the Toeplitz quantization map $\CCT_n$ to be a Lie algebra homomorphism is of order $\CO(1/n)$, in the limit, the deviation of $T^i_{(n)}(s)$ from $\CCT_n(t^i(s))$ vanishes:
\begin{equation}
 \int_{\CI^{(n)}_2}\dd s~ ||T^i_{(n)}(s)-\CCT_n(t^i(s))||^2\ \rightarrow \ 0~~~\mbox{as}~~~n\rightarrow \infty~.
\end{equation}
Here, $\CI^{(n)}_2$ is the maximal interval on which both the $T^i_{(n)}(s)$ and the $t^i(s)$ are defined. As the functions $t^i(s)$ diverge at $s=v$, the same should hold for the matrix-valued functions $T^i_{(n)}$ in a neighborhood of $v$ that becomes smaller with $n$, i.e.\ $\CI^{(n)}_2\rightarrow \CI_2$. The $T^i_{(n)}(s)$ thus indeed describe a sequence of Nahm data that encodes monopole solution and converges to the $\Pi_\omega$-Nahm data $t^i(s)$ of a magnetic bag. These arguments suggest that Bolognesi's conjecture is true at least for degenerate bags like e.g.\ magnetic discs.

For non-degenerate bags, the situation is more subtle: the extension of the $\Pi_\omega$-Nahm data to $\CI_2$ via $t^i(x,-s):=t^i(Rx,s)$ has a discontinuity at $s=0$ as $t^i(Rx,0)\neq t^i(x,0)$ for some $i=1,2,3$. Thus the limiting configuration $t^i(s)$ must satisfy a modified Nahm equation.  Because the $t^i(x,s)$ satisfy the Nahm equation on $\CI_2\backslash\{0\}$, we are led to
\begin{equation}\label{eq:NahmJumpBolognesi}
 \dderr{t^i}{s}=\frac{2\pi}{q}\eps_{ijk}\{t^j,t^k\}+\zeta^i \delta(s)~,
\end{equation}
where $\zeta^i\in\CC^\infty(S^2)$ determines the size of the jump at $s=0$.  Solutions of this modified Nahm equation are expected to be good approximations to solutions of the usual Nahm equation in the large $n$ limit.  To understand this modification in more detail, let us turn to the brane interpretation of magnetic domains in string theory.

\section{Brane interpretation}\label{sec:branes}

\subsection{Brane interpretation of magnetic walls and bags}\label{dbraneint}

In string theory, BPS monopoles can be interpreted as D1-branes ending on D3-branes \cite{Diaconescu:1996rk,Tsimpis:1998zh}. Explicitly, we consider the following configuration of $n$ D1-branes ending on $N$ D3-branes at positions $x^6=s_i$, $i=1,\ldots,N$:
\begin{equation}\label{diag:D1D3}
\begin{tabular}{rcccccccc}
& 0 & 1 & 2 & 3 & 4 & 5 & 6 & \ldots\\
D1 & $\times$ & & & & & & $\vdash$ \\
D3 & $\times$ & $\times$ & $\times$ & $\times$ & & & $s_i$
\end{tabular}
\end{equation}
A $\times$ indicates a direction that is fully contained in the brane's worldvolume, while a $\vdash$ indicates that the brane's worldvolume is bounded in this direction. To compare with Chalmers-Hanany-Witten configurations \cite{Chalmers:1996xh,Hanany:1996ie}, we T-dualize along the $x^4$- and $x^5$-direc\-tions, S-dualize and obtain
\begin{equation}\label{diag:D3NS5}
\begin{tabular}{rcccccccc}
& 0 & 1 & 2 & 3 & 4 & 5 & 6 & \ldots\\
D3 & $\times$ & & & & $\times$ & $\times$ & $\vdash$ \\
NS5 & $\times$ & $\times$ & $\times$ & $\times$ & $\times$ & $\times$ & $s_i$
\end{tabular}
\end{equation}
Dirac monopoles correspond to a single $N=1$ NS5-brane at e.g.\ $s=0$. The usual $\sSU(2)$-monopoles yield $N=2$ NS5-branes at positions $s_1=-v$ and $s_2=v$ and D3-branes suspended between them, where $\CI=(-v,v)$ is the interval over which the Nahm data is supported. The BPS equations in the gauge theory description of configuration \eqref{diag:D3NS5} are just the ordinary Nahm equations, cf.\ e.g.\ \cite{Hanany:1996ie,Cherkis:2008ip,Cherkis:2011ee}. 	

As a first nontrivial configuration, let us consider a so-called monopole wall \cite{Ward:2006wt}, i.e.\ a doubly periodic monopole. A brane interpretation of such a monopole wall has been recently discussed in \cite{Cherkis:2012qs}. Here,  we consider an NS5-brane at $s=0$ and D3-branes whose endpoints form a two-dimensional lattice in the $\FR^2_{12}$-directions\footnote{Subscripts on manifolds denote the directions in which these spaces extend into the target space $\FR^{1,9}$.}. Alternatively, we can replace the subspace $\FR^3_{123}$  with $T^2_{12}\times \FR_3$ and consider a single monopole on this space at $x^1=x^2=0$. Let us assume that the radii of the torus $T^2_{12}$ are sufficiently small and therefore the Higgs field $\Phi$ is effectively constant in the compactified directions. It therefore satisfies the Laplace equation in the $x^3$-direction:
\begin{equation}
 \der{x^3}\der{x^3} \phi(x_3)=\tan\theta~\delta(0)~,
\end{equation}
where $x^3=0$ is the position of the endpoint of the D3-brane on the NS5-brane and the angle $\theta$ is related to the lattice spacing or, equivalently, the radii of the torus $T^2_{12}$, cf.\ e.g.\ \cite{Aharony:1997ju}. The solution of this equation is
\begin{equation}
 \phi(x_3)=\frac{\tan\theta}{2}|x^3|+b x^3+c~,~~~b,c\in\FR~.
\end{equation}
The constants can be fixed by demanding that $x^6=\phi(x^3)=0$ for $x^3\leq 0$, which yields $b=\tfrac{\tan\theta}{2}$ and $c=0$. This configuration is in fact related to a bound state between D5- and NS5-branes. To see this, let us T-dualize along $T^2_{12}$, and we arrive at the configuration
\begin{equation*}\label{diag:pq}
\begin{tabular}{rcccccccc}
& 0 & 1 & 2 & 3 & 4 & 5 & 6 & \ldots\\
D5 & $\times$ & $\times$ & $\times$ & 0 & $\times$ & $\times$ & $\vdash$ \\
NS5 & $\times$ & $\times$ & $\times$ & $\vdash$ & $\times$ & $\times$ & 0\\
(1,1) & $\times$ & $\times$ & $\times$ & \rotatebox[origin=c]{45}{$\vdash$} & $\times$ & $\times$ &  \rotatebox[origin=c]{45}{$\vdash$}
\end{tabular}
\begin{picture}(290,55)(0,0)
\put(30.0,20.0){\line(1,0){50}}
\put(40.0,12.0){\makebox(0,0)[c]{NS5}}
\put(80.0,20.0){\line(0,-1){50}}
\put(90.0,-22.0){\makebox(0,0)[c]{D5}}
\put(80.0,20.0){\line(1,1){25}}
\put(130.0,36.0){\makebox(0,0)[c]{$(1,1)$-brane}}
\put(130.0,-25.0){\vector(1,0){20}}
\put(130.0,-25.0){\vector(0,1){20}}
\put(160.0,-25.0){\makebox(0,0)[c]{$x^3$}}
\put(130.0,2.0){\makebox(0,0)[c]{$x^6$}}
\end{picture}\vspace*{0.4cm}
\end{equation*}
The NS5-brane ends at $x^3=0$ and turns into a $(p,q)$-fivebrane with $p=q=1$ which extends diagonally in $\FR^2_{36}$ as indicated by the symbol \rotatebox[origin=c]{45}{$\vdash$}. A $(p,q)$-brane \cite{Aharony:1997ju} is a bound state of $p$ NS5-branes and $q$ D5-branes, fused together at a junction like the one above. The angle $\theta$ is restricted by $\tan\theta=g_s \frac{p}{q}$, $p,q\in\NN$, where $g_s$ is the string coupling.

Combining two such monopole walls and tuning the length of the connecting D5-branes to zero, we obtain the following picture:
\begin{equation*}
\begin{picture}(130,55)
\put(30.0,20.0){\line(1,0){50}}
\put(40.0,12.0){\makebox(0,0)[c]{NS5}}
\put(30.0,-10.0){\line(1,0){50}}
\put(40.0,-18.0){\makebox(0,0)[c]{NS5}}
\put(80.0,20.0){\line(0,-1){30}}
\put(90.0,6.0){\makebox(0,0)[c]{D5}}
\put(80.0,20.0){\line(1,1){25}}
\put(130.0,36.0){\makebox(0,0)[c]{$(1,1)$-brane}}
\put(80.0,-10.0){\line(1,-1){25}}
\put(130.0,-26.0){\makebox(0,0)[c]{$(1,1)$-brane}}
\end{picture}~~~~\longrightarrow
\begin{picture}(200,45)
\put(30.0,6.0){\line(1,0){50}}
\put(30.0,4.0){\line(1,0){50}}
\put(40.0,-5.0){\makebox(0,0)[c]{NS5}}
\put(80.0,6.0){\line(0,-1){2}}
\put(80.0,6.0){\line(1,1){25}}
\put(130.0,38.0){\makebox(0,0)[c]{$(1,1)$-brane}}
\put(80.0,4.0){\line(1,-1){25}}
\put(130.0,-28.0){\makebox(0,0)[c]{$(1,1)$-brane}}
\end{picture}\vspace*{1.2cm}
\end{equation*}
The right configuration is a useful picture for the neighborhood of the edge of a magnetic bag. Let us now try to model a complete spherical magnetic bag. For this, consider two NS5-branes as above, which extend into $\FR^6_{012345}$ at $s_1=-v$ and $s_2=v$, together with D5-branes extending into $\FR_0\times \FR^3_{456}$, wrapping a 2-sphere $S^2$ in $\FR^3_{123}$, and ending on the NS5-branes at $s_1$ and $s_2$. The boundary of the D5-branes in $\FR^3_{123}$ is given by the 2-sphere, which is identified with the surface of the magnetic bag. We now perform again the analysis of the Higgs field as above. The Higgs field now has to satisfy the Laplace equation in three dimensions, which yields $\phi\sim v-\frac{1}{r}$, where $r$ is the radial distance from the center of the 2-spheres $S^2$. As a boundary condition, we demand that the NS5-branes are flat in the interior of the bag. This deforms them to $(1,1)$-branes on the outside of the bag:
\begin{equation}\label{diag:pq2}
\begin{tabular}{rc}
D5 & $\FR_0\times S^2_{123}\times \FR^3_{456}$\\ 
NS5 & $\FR_0\times B^3_{123}\times \FR^2_{45}$\\ 
(1,1) & $\FR_0\times S^2_{123}\times \FR^+_{1236}\times \FR^2_{45}$
\end{tabular}
\begin{picture}(190,35)
\put(15.0,15.0){\line(1,1){10}}
\put(165.0,15.0){\line(1,1){10}}
\put(15.0,-25.0){\line(1,1){10}}
\put(165.0,-25.0){\line(1,1){10}}
\put(15.0,15.0){\line(1,0){150}}
\put(25.0,25.0){\line(1,0){150}}
\put(25.0,25.0){\line(1,0){150}}
\put(15.0,-25.0){\line(1,0){150}}
\put(25.0,-15.0){\line(1,0){150}}
\qbezier(95, -10)(120, -10)(120, 0)
\qbezier(95, 10)(120, 10)(120, 0)
\qbezier(95, 10)(70, 10)(70, 0)
\qbezier(95, -10)(70, -10)(70, 0)
\qbezier(120, 0)(120, 20)(170, 20)
\qbezier(120, 0)(120, -20)(170, -20)
\qbezier(70, 0)(70, -20)(20, -20)
\qbezier(70, 0)(70, 20)(20, 20)
\put(95.0,0.0){\makebox(0,0)[c]{NS5}}
\put(190.0,20.0){\makebox(0,0)[c]{$(1,1)$}}
\put(190.0,-20.0){\makebox(0,0)[c]{$(1,1)$}}
\end{picture}\vspace*{0.6cm}
\end{equation}
After taking the length of the D5-branes in the $x^6$ direction to zero, the Higgs field has the profile of that of the spherical magnetic bag.

\subsection{Approximating the Nahm data for magnetic bags}

We now return to equation \eqref{eq:NahmJumpBolognesi} and its interpretation in terms of branes. We start again from two NS5 branes at $s_1=-v$ and $s_2=v$ and $n$ D3-branes suspended between them. The source at $s=0$ in equation \eqref{eq:NahmJumpBolognesi} signals that there is an `impurity' in the worldvolume of the D3-branes. Such impurity theories have been extensively studied, see e.g.\ \cite{Cherkis:2012qs} and references therein. In string theory, the impurities can be modeled by inserting fivebranes whose worldvolumes are orthogonal to the direction $x^6$. These fivebranes are assumed to be heavy compared to the D3-branes, and therefore they are considered as static. Moreover, the distribution-like source induces a jump in the Nahm datum $T^1$ at $s=0$, signaling a breaking of the D3-branes in the $x^1$-direction:
\begin{equation}\label{diag:pq3}
\begin{picture}(170,60)
\put(0.0,0.0){\line(0,1){50}}
\put(0.0,10.0){\line(1,0){50}}
\put(0.0,12.0){\line(1,0){50}}
\linethickness{1pt}
\put(50.0,0.0){\line(0,1){50}}
\linethickness{0.4pt}
\put(50.0,40.0){\line(1,0){50}}
\put(50.0,38.0){\line(1,0){50}}
\put(100.0,0.0){\line(0,1){50}}
\put(13.0,45.0){\makebox(0,0)[c]{NS5}}
\put(113.0,45.0){\makebox(0,0)[c]{NS5}}
\put(66.0,5.0){\makebox(0,0)[c]{defect}}
\put(25.0,17.0){\makebox(0,0)[c]{D3s}}
\put(75.0,32.0){\makebox(0,0)[c]{D3s}}
\put(140.0,10.0){\vector(1,0){20}}
\put(140.0,10.0){\vector(0,1){20}}
\put(170.0,10.0){\makebox(0,0)[c]{$x^6$}}
\put(140.0,37.0){\makebox(0,0)[c]{$x^1$}}
\end{picture}\vspace*{0.0cm}
\end{equation}
If we insert a D5-brane at $s=0$ parallel to the NS5-branes and such that the D3-branes can intersect it, the strings connecting the D3- and D5-branes yield an additional fundamental hypermultiplet \cite{Hanany:1996ie}. Giving a vacuum expectation value (vev) to this hypermultiplet, we obtain additional source terms to the Nahm equation, which are of the form\footnote{As remarked in \cite{Tsimpis:1998zh}, it is expected that stringy effects will regulate the $\delta(s)$-term to an exponential approximation.}
\begin{equation}\label{eq:NahmJumpFinite}
 \dderr{T^i_{(n)}}{s}=\frac{2\pi}{q}\eps_{ijk}[T^j_{(n)},T^k_{(n)}]+h_a\otimes h_b^*\sigma^i_{ab}\delta(s)~,
\end{equation}
where $h_a\in \FC^k\otimes \FC^2$. This is the Nahm equation appearing in the construction of an $\sSU(3)$ monopole \cite{Hurtubise:1989qy,Tsimpis:1998zh}. Note that the expression $h_a\otimes h_b^*\in\sEnd(\FC^2)\otimes \sEnd(\FC^k)$ is of rank one in the gauge part $\sEnd(\FC^k)$. This amounts to the fact that only one of the D3-branes suspended between the two NS5-branes can break up on the D5-brane\footnote{This is also related to the s-rule \cite{Hanany:1996ie}, which states that only one D3-brane can be supersymmetrically suspended between any given pair of NS5- and D5-branes.}. Here, however, we want all the D3-branes to break in the $x^1$ direction. 

The alternative is to insert an NS5-brane. This generates an additional bifundamental hypermultiplet at $s=0$ arising from strings connecting the D3-branes to the left and the right of the NS5-brane \cite{Hanany:1996ie,Cherkis:2012qs}. Giving a vev to this hypermultiplet, we obtain the Nahm equation
\begin{equation}\label{eq:NahmJumpFinite2}
 \dderr{T^i_{(n)}}{s}=\frac{2\pi}{q}\eps_{ijk}[T^j_{(n)},T^k_{(n)}]+\zeta^i\delta(s)~,
\end{equation}
which is the finite $n$ version of \eqref{eq:NahmJumpBolognesi}. Here, $\zeta^i\in\au(n)$ is determined by the vev of the hypermultiplet. This configuration, however, does not describe an $\sSU(2)$ monopole. In fact, the configuration we arrived at is S-dual to a sequence of D5-branes at $s=-v$, $s=0$ and $s=v$, which is the usual description of an $\sSU(3)$ monopole, except for the fact that all the D3-branes break on the D-brane in the middle. To obtain a brane configuration corresponding to an $\sSU(2)$-monopole, we compactify the direction $x^6$ on a circle and identify the NS5-branes at $s=-v$ and $s=+v$. On the latter NS5-brane, the D3-branes end with the usual Nahm boundary condition, 
while on the NS5-brane at $s=0$, they break up and their worldvolume becomes discontinuous in the $x^1$-direction. Inverting the process of T- and S-dualizing, we recover a D-brane configuration with two D3-branes and $2n$ D1-branes, which describes an $\sSU(2)$-monopole configuration. While this string theory interpretation is certainly no proof of the Bolognesi conjecture, it gives at least strong evidence for its validity.

Finally, let us try to connect configuration \eqref{diag:pq2} to \eqref{diag:pq3}. While \eqref{diag:pq3} is the na\"ive, classical picture, configuration \eqref{diag:pq2} incorporates quantum corrections bending the branes. We know that each point of the worldvolume of the D3-branes in \eqref{diag:pq3} polarizes into a fuzzy sphere due to the Myers effect \cite{Myers:1999ps,Constable:1999ac}. In the limit $n\rightarrow \infty$, the D3-branes therefore turn into D5-branes wrapping a sphere $S^2_{123}$. Moreover, if we assume that all the D3-branes come in pairs such that the configuration \eqref{diag:pq3} is symmetric with respect to the $x^6$-coordinate axes, we arrive at the following quantum corrected picture:
\begin{equation*}
\begin{picture}(190,60)
\put(0.0,10.0){\line(-1,-1){15}}
\put(0.0,40.0){\line(-1,1){15}}
\put(100.0,10.0){\line(1,-1){15}}
\put(100.0,40.0){\line(1,1){15}}
\put(0.0,10.0){\line(0,1){30}}
\put(0.0,10.0){\line(1,0){50}}
\put(0.0,12.0){\line(1,0){50}}
\linethickness{1pt}
\put(50.0,0.0){\line(0,1){50}}
\linethickness{0.4pt}
\put(50.0,40.0){\line(1,0){50}}
\put(50.0,38.0){\line(1,0){50}}
\put(0.0,40.0){\line(1,0){50}}
\put(0.0,38.0){\line(1,0){50}}
\put(50.0,10.0){\line(1,0){50}}
\put(50.0,12.0){\line(1,0){50}}
\put(100.0,10.0){\line(0,1){30}}
\put(-23.0,45.0){\makebox(0,0)[c]{$(p,q)$}}
\put(-23.0,5.0){\makebox(0,0)[c]{$(p,q)$}}
\put(125.0,45.0){\makebox(0,0)[c]{$(p,q)$}}
\put(125.0,5.0){\makebox(0,0)[c]{$(p,q)$}}
\put(-13.0,25.0){\makebox(0,0)[c]{NS5}}
\put(113.0,25.0){\makebox(0,0)[c]{NS5}}
\put(66.0,3.0){\makebox(0,0)[c]{defect}}
\put(25.0,17.0){\makebox(0,0)[c]{D5s}}
\put(75.0,32.0){\makebox(0,0)[c]{D5s}}
\put(75.0,17.0){\makebox(0,0)[c]{D5s}}
\put(25.0,32.0){\makebox(0,0)[c]{D5s}}
\put(160.0,10.0){\vector(1,0){20}}
\put(160.0,10.0){\vector(0,1){20}}
\put(190.0,10.0){\makebox(0,0)[c]{$x^6$}}
\put(160.0,37.0){\makebox(0,0)[c]{$x^1$}}
\end{picture}\vspace*{0.2cm}
\end{equation*}
Up to the defect at $s=0$, this configuration is identical to \eqref{diag:pq2}. Note that to obtain a magnetic bag, we have to tune the distance between the NS5-branes to zero. To our knowledge, it is still unclear how to describe Chalmers-Hanany-Witten configurations with stacks of multiple NS5-branes as impurities. Studying our example of a magnetic bag in more detail might provide some new insights into this issue. In particular, it might explain the appearance of the additional defect at $s=0$. These considerations, however, are beyond the scope of this paper.

It is clear that the D-brane configurations we considered in this section all have lifts to M-theory. In particular, the M-brane configuration obtained from lifting \eqref{diag:pq2} describes a bag of self-dual strings, which are bounded by three-dimensional surfaces diffeomorphic to $S^3$. We will present the corresponding Nahm constructions in the following.

\section{Magnetic domains in four dimensions}\label{sect:SDS}

\subsection{From self-dual strings to magnetic domains}

A self-dual string is a BPS configuration of M2-branes ending on M5-branes in the following way:
\begin{equation}\label{diag:M2M5}
\begin{tabular}{rccccccc}
${\rm M}$ & 0 & 1 & 2 & 3 & 4 & 5 & 6 \\
M2 & $\times$ & & & & & $\times$ & $\vdash$ \\
M5 & $\times$ & $\times$ & $\times$ & $\times$ & $\times$ & $\times$ 
\end{tabular}
\end{equation}
This configuration is obtained from \eqref{diag:D1D3} by T-dualizing along the $x^5$-direction and interpreting the $x^4$-direction as the M-theory direction.

If only one M5-brane is present, the effective description on the M5-brane consists of a $\au(1)$-valued scalar field $\phi$ and a closed $\au(1)$-valued 3-form $h$ on $\FR^4$.  These must satisfy the \emph{self-dual string equation} \cite{Howe:1997ue},
\begin{equation}\label{eq:SDS}
h=*\dd\phi~,
\end{equation}
which is the M-theory analog of the Bogomolny equation for Dirac monopoles.  For two or more M5-branes the effective description on the M5-brane worldvolume is not known. There are some suggestions using loop space, and we will return to these in section \ref{sect:LSSDS}.

We saw in sections \ref{sec:Bags} and \ref{sec:branes} that magnetic domains, obeying an abelian equation, can appear in the $n\to\infty$ limit of $n$ D1-branes stretched between two D3-branes. We expect something similar to happen here: If we consider the limit of infinitely many M2-branes stretched between two M5-branes, the theory should become abelian. We will refer to the resulting configurations again as {\em magnetic domains}. These domains in four dimensions are described by a Higgs field $\phi$ and a closed 3-form $h$ in a domain $\Omega\subset\FR^4$, both taking values in $\au(1)$ and having the following properties:
\begin{itemize}
 \setlength{\itemsep}{-1mm}
 \item $h$ is closed, and therefore we have locally a 2-form potential $b$ with $h=\dd b$,
 \item $h$ and $\phi$ satisfy the self-dual string equation $h=*\dd \phi$ in the region $\Omega\subset \FR^4$,
 \item $\dd \phi\neq 0$ in $\Omega$ and
 \item depending on the shape and dimensionality of the boundary of the domain $\Omega$, $\phi$ satisfies certain boundary conditions.
\end{itemize}
We clearly expect to find four-dimensional generalizations of the magnetic domains we know from three dimensions, in particular magnetic bags, magnetic tubes and magnetic walls. Analogously to the name monopole bags, we will refer to magnetic bags in four dimensions as self-dual string bags.

It is interesting to note that, similar to the Yang-Mills-Higgs energy functional, we can define a functional
\begin{equation}
E=\tfrac{1}{2}\int_\Omega h\wedge *h+\dd\phi\wedge *\dd\phi~,
\end{equation}
which has a Bogomolny bound 
\begin{equation}
E=\int_\Omega \tfrac{1}{2}|\dd\phi-*h|^2+d\phi\wedge h~\ge vq~,~~~q:=\int_{S^3_\infty}h~,
\end{equation}
saturated by solutions to the self-dual string equation \eqref{eq:SDS}. 

\subsection{Nambu-Poisson structure and the Basu-Harvey equation}

To develop the Nahm construction for magnetic domains in four dimensions, we first have to review the corresponding Nahm equation. In the case of self-dual strings, this is the Basu-Harvey equation \cite{Basu:2004ed}
\begin{equation}\label{eq:BasuFinite}
\frac{\dd T^\mu}{\dd s} = \frac{e}{3!}\,\eps_{\mu\nu\rho\sigma}[T^\nu,T^\rho,T^\sigma]~.
\end{equation}
The functions $T^\mu(s)$ take values in a {\em 3-Lie algebra}. 3-Lie algebras were introduced by Filippov \cite{Filippov:1985aa} and are by definition vector spaces $\CA$ equipped with a 3-bracket $[\cdot,\cdot,\cdot]:\CA^{3}\rightarrow \CA$.  The 3-bracket is linear and anti-symmetric in all of its arguments, and is required to satisfy the fundamental identity,
\begin{equation}\label{eq:FI3Alg}
\begin{aligned}
{}[f_1,f_2,[g_1,g_2,g_3]]&=[[f_1,f_2,g_1],g_2,g_3]+[g_1,[f_1,f_2,g_2],g_3]+[g_1,g_2,[f_1,f_2,g_3]] ~.
\end{aligned}
\end{equation}

The most prominent example of a 3-Lie algebra is the 3-Lie algebra called $\CA_4$ in the classification of \cite{Filippov:1985aa}.  As a vector space, $\CA_4$ is isomorphic to $\FR^4$, and its generators $\tau_1,\dots,\tau_4$ satisfy
\begin{equation}
 [\tau_\mu,\tau_\nu,\tau_\rho]=\eps_{\mu\nu\rho\sigma}\tau_\sigma~,~~~\mu\nu\rho\sigma=1,\ldots,4~.
\end{equation}
This 3-Lie algebra (and its direct sums) is known to be the only non-trivial, finite-dimensio\-nal example of a 3-Lie algebra endowed with a positive definite invariant inner product \cite{Nagy:2007aa}.

In contrast, there are many examples of infinite-dimensional normed 3-Lie algebras.  Let $M$ be any 3-manifold equipped with a non-vanishing volume form $\omega$.  The space $\CC^{\infty}(M)$ of smooth functions forms a 3-Lie algebra, with 3-bracket defined by the equation
\begin{equation}
 \{f,g,h\}\omega = \dd f\wedge \dd g\wedge\dd h~.
\end{equation}
In addition to the fundamental identity \eqref{eq:FI3Alg}, the 3-bracket satisfies the Leibniz rule
\begin{equation}\label{eq:Leibniz}
 \{f_1f_2,g,h\} = f_1\{f_2,g,h\} + \{f_1,g,h\}f_2 ~.
\end{equation}
This implies that for any $g,h\in \CC^\infty(M)$ the map $D(g,h):f\to\{g,h,f\}$ is a derivation, which means that $D(g,h)$ is a vector field.  In general, a 3-Lie algebra structure on the algebra of functions over a manifold obeying the Leibniz rule is called a {\em Nambu-Poisson structure} \cite{Nambu:1973qe,Takhtajan:1993vr}.

Solutions to the Basu-Harvey equation based on the 3-Lie algebra $\CA_4$ are conjectured to describe two M2-branes stretching between M5-branes.  We will show below that the appropriate 3-Lie algebra for describing self-dual string bags is $\CC^\infty(S^3)$ equipped with the Nambu-Poisson 3-Lie bracket induced by the $\sSO(4)$-invariant volume form $\omega$.  In standard polar coordinates $0\le\theta^1,\theta^2\le\pi$,  $0\le\theta^3\le2\pi$ the volume form is
\begin{equation}
\omega=\sin^2\theta^1 \sin \theta^2~\dd\theta^1\wedge\dd\theta^2\wedge\dd\theta^3~,
\end{equation}
and the Nambu-Poisson 3-bracket is given by
\begin{equation}
\{f,g,h\}=\frac{1}{\sin^2\theta^1 \sin \theta^2} \eps^{ijk}\frac{\pa f}{\pa \theta^{i}}\frac{\pa g}{\pa \theta^j}\frac{\pa h}{\pa \theta^{k}}~~.
\end{equation}
It will be convenient to denote by $x^1,x^2,x^3,x^4$ the functions on $S^3$ obtained by restricting coordinate functions from $\FR^4$.  These of course satisfy $x^\mu x^\mu=1$, and their 3-brackets with each other are
\begin{equation}
 \{ x^\mu, x^\nu, x^\rho \} = \eps^{\mu\nu\rho\sigma} x^\sigma ~.
\end{equation}
Thus the $x^\mu$ span a sub-algebra of $\CC^\infty(S^3)$ isomorphic to $\CA_4$.

\subsection{Nahm transform and its inverse}\label{sect:SDStransform}

In general, we will denote the Nambu-Poisson structure on a three-dimensional manifold $M$ induced by its volume form $\omega$ by $\Pi_\omega$. Under {\em $\Pi_\omega$-Basu-Harvey data} for magnetic domains in four dimensions, we understand a set of four functions $t^\mu$ on $M\times \CI$, where $\CI$ is a union of finitely many intervals, satisfying the $\Pi_\omega$-Basu-Harvey equation 
\begin{equation}\label{eq:BHInfinite}
\frac{\dd t^{\mu}}{\dd s}=\frac{2\pi^2}{ 3! q}~\eps^{\mu\nu\kappa\lambda}\{t^{\nu},t^{\kappa},t^{\lambda}\}_\omega ~.
\end{equation}
We will discuss in section \ref{sect:HermitianBH} how this Basu-Harvey equation emerges as the large $n$ limit of Basu-Harvey equations based on hermitian 3-algebras. Analogously to the case of magnetic domains in $\FR^3$, we have the following theorem, which refines a result of Dunajski \cite{Dunajski:2003ep}: 

\begin{thm}
\label{thm:4}
Up to gauge equivalence, there is a one-to-one correspondence between
\vspace{-0.2cm}
\begin{itemize}
 \setlength{\itemsep}{-1mm}
 \item sets of $\Pi_\omega$-Basu-Harvey data with the property that the map from $M\times \CI$ to $\Omega\subset\FR^4$ defined by the $t^\mu$ is a diffeomorphism $t:M\times \CI\rightarrow \Omega$, and
 \item magnetic domains $\Omega$ that are diffeomorphic to $M\times \CI$, where the restriction of the 3-form curvature $h$ to any slice $M\times\{s_0\}$ has the same volume type as $\omega$ and $\CI$ is the range of $\phi$. Explicitly, there is a diffeomorphism $u:\Omega\rightarrow M\times \CI$ such that $h=\frac{q}{2\pi^2}u^*\omega$ and $\phi=s\circ u$ on $\Omega$. 
\end{itemize}
\end{thm}

The proof is a minor generalization of that of theorem \ref{thm:2}. The Basu-Harvey data defines a diffeomorphism $t$ from $M\times \CI$ to a subset $\Omega\subset\FR^4$ with inverse $u$.  By definition of the Nambu 3-bracket, the infinite-charge Basu-Harvey equation \eqref{eq:BHInfinite} is equivalent to
 \begin{equation}
\dd t^{\mu}\wedge \omega=\frac{2\pi^2}{3! q}~\eps_{\mu\nu\kappa\lambda}\dd t^{\nu}\wedge\dd t^{\kappa}\wedge \dd t^{\lambda}\wedge\dd s~.
\end{equation}
This implies the following equation on $\FR^4$:
\begin{equation}
 \dd y^{\mu}\wedge u^*\omega=\frac{2\pi^2}{q}\dd y^\mu\wedge*\dd (u^* s)~.
\end{equation}
Thus $\phi=u^* s$ and $h=\frac{q }{2\pi^2}~u^*\omega$ solve the self-dual string equation \eqref{eq:SDS}.

To define the inverse transform we restrict ourselves again to a connected component. We choose a value $\phi_0=s_0\in\CI$, which yields the level surface $\Sigma_{\phi_0}$, which is embedded in $\Omega$ via the map $i:\Sigma_{\phi_0}\embd \Omega$. Because the restriction of $h$ and $\omega$ have the same volume type, there is a diffeomorphism $w:M\rightarrow \Sigma_{\phi_0}$ such that $w^*i^*h=\frac{q}{2\pi^2}\omega$. The diffeomorphism $w\circ i$ can be extended to all of $M\times \CI$ by solving the differential equation 
\begin{equation}
 \frac{\dd y^\mu}{\dd s} = Y^\mu(y(s))~,~~~Y=\left(\derr{\phi}{y^\nu}\derr{\phi}{y_\nu}\right)^{-1}\derr{\phi}{y^\mu}\der{y_\mu}
\end{equation}
with boundary condition $y(s_0)=w\circ i$. Here, the solution $y$ can again be identified with the diffeomorphism $t:M\times \CI\rightarrow \Omega$. It can be readily checked that this construction inverts the Nahm transform.

\subsection{Examples}

First we consider what we will call {\em self-dual string bags}: magnetic domains in four dimensions for which $\Omega$ is the exterior of a hypersurface $\Sigma\subset\FR^4$ diffeomorphic to $S^3$. On the interior of this hypersurface $\Sigma$, we have $\phi=0$, and on the exterior, $\phi\sim v-\frac{1}{r^2}$ as $r\rightarrow \infty$.  The corresponding $\Pi_\omega$-Basu-Harvey data consists of four functions on $S^3\times [0,v)$ satisfying
\begin{equation}
t^{\mu}\sim\frac{ x^\mu}{2\pi}\left(\frac{q}{v-s}\right)^\frac{1}{2}~\mbox{ as }s\rightarrow v~.
\end{equation}

The simplest example of $\Pi_\omega$-Basu-Harvey data is
\begin{equation}\label{nahmboundary}
t^{\mu}=\frac{ x^\mu}{2\pi}\left(\frac{q}{v-s}\right)^\frac{1}{2}~.
\end{equation}
The image of the map $t:S^3\times [0,v)\to\FR^4$ is the set $\Omega=\{r^2\geq q/4\pi^2v\}$, and the inverse map $u:\Omega\rightarrow S^3\times \CI$ is 
\begin{equation}
u(\vec y)=\left(\frac{\vec y}{r},v-\frac{q}{4\pi^2 r^2}\right).
\end{equation}
Thus the corresponding self-dual string bag is the following spherically-symmetry configuration:
\begin{equation}
\begin{aligned}
\phi =& \begin{cases} v-\frac{q}{4\pi^2 r^2} & r^2\geq \frac{q}{4\pi^2v} \\ 0 & r^2 < \frac{q}{4\pi^2v} \end{cases}~,\\
h =& \begin{cases} \frac{q}{3!2\pi^2r^4} \eps_{\mu\nu\rho\sigma}y^\mu\dd y^\nu\wedge\dd y^\rho\wedge\dd y^\sigma & r^2\geq \frac{q}{4\pi^2v} \\ 0 & r^2 < \frac{q}{4\pi^2v} \end{cases} ~.
\end{aligned}
\end{equation}

Another example of Basu-Harvey data\footnote{The corresponding solution to the Basu-Harvey equation \eqref{eq:BasuFinite} with Nahm data taking values in the 3-Lie algebra $A_4$ was given in \cite{Palmer:2011vx}.}, this time describing an ellipsoidal bag extended in the $y^4$ direction, is given by
\begin{equation}
t\left( x^\mu,s\right)= \sqrt \frac{q}{2\pi^2} \left(\frac{ x^{i}}{\sqrt{(v-s)(2+v-s)}},\frac{ x^{4}(1+v-s)}{\sqrt{(v-s)(2+v-s)}} \right)~,~~i=1,\dots,3~.
\end{equation}
The inverse map $u:\Omega\rightarrow S^3\times \CI$ is then
\begin{equation}
u(y)=\left(\left(\frac{y^i}{\alpha},\frac{y^4}{\alpha\sqrt{1+\frac{q}{2\pi^2\alpha^2}}}\right),v+1-\sqrt{1+\frac{q}{2\pi^2\alpha^2}}\right)~,
\end{equation}
where 
\begin{equation}
\alpha^2=\tfrac{1}{2}\left(r^2-\frac{q}{2\pi^2}+\sqrt{\frac{2q}{\pi^2}(r^2-(y^4)^2)+(r^2-\frac{q}{2\pi^2})^2}\right)~.
\end{equation}
This gives the magnetic domain
\begin{equation}
\begin{aligned}
&\phi=v+1-\sqrt{1+\frac{q}{2\pi^2\alpha^2}}~,~~ h=\eps_{\mu\nu\rho\sigma} \frac{q}{2\pi^2\alpha^3\sqrt{1+\frac{q}{2\pi^2\alpha^2}}}\frac{\pa\alpha}{\pa y^\mu}\dd y^\nu\wedge\dd y^\rho\wedge\dd y^\sigma~\\
&\mbox{on}~~~ \Omega = \left\{ y~\Big|~(y^1)^2+(y^2)^2+(y^3)^2+\frac{1}{(1+ v)^2} (y^4)^2\geq \frac{q}{2\pi^2(2v+v^2)}\right\}\subset \FR^4~.
\end{aligned}
\end{equation}

Analogously to the 3-dimensional examples presented in section \ref{sec:3DExamples}, one can also construct 4-dimensional magnetic domains from manifolds $M=\FR^3$, $M=\FR^2\times S^1$ and $M=\FR\times S^2$ endowed with a volume form. The boundary conditions for the scalar field $\phi$ can be fixed by demanding that $\phi$ asymptotes to a harmonic function with appropriate symmetries, and these induce boundary conditions on the Basu-Harvey data.

\subsection{Conserved charges}\label{sect:cons}

It is interesting to note that the Basu-Harvey equation is integrable.  Rather than a Lax pair, the integrability manifests itself through a Lax triple $(t,A,B)$ with spectral parameters $\eta,\zeta\in\CPP^1$:
\begin{equation}
\begin{aligned}
 t(\eta,\zeta) &= (t^1+\mathrm{i} t^2) + \zeta (t^3+\mathrm{i} t^4) + \eta (t^3-\mathrm{i} t^4) + \zeta\eta (-t^1+\mathrm{i} t^2)~, \\
 A(\eta) &=  (t^3+\mathrm{i} t^4) + \eta (-t^1+\mathrm{i} t^2)~, \\
 B(\zeta) &=  (t^3-\mathrm{i} t^4) + \zeta (-t^1+\mathrm{i} t^2)~.
\end{aligned}
\end{equation}
Here, $t^i\in \CC^\infty(\CI)\otimes \CA$, where $\CA$ is a 3-Lie algebra. Using the anti-symmetry of the 3-bracket, it can be shown that the Basu-Harvey equation is equivalent to
\begin{equation}
\label{Lax equation}
 \frac{\mathrm{d}}{\mathrm{d} s} t(\eta,\zeta) = [A(\eta),B(\zeta),t(\eta,\zeta)]~.
\end{equation}

Specializing now to the 3-Lie algebra $\CC^\infty(M)$, we define an infinite tower of conserved charges by taking the coefficients of the polynomials,
\begin{equation}
\int_{M\times s_0} \omega~ t(\eta,\zeta)^n~, \quad n\in\NN~,
\end{equation}
where $\omega$ is again the volume form on $M$.  We assume that these integrals converge, which is certainly the case when $M$ is compact.  The fact that these quantities are conserved follows from the Lax equation \eqref{Lax equation} and the observation that the integral of the 3-bracket of any three functions is zero.

The conserved charges can equivalently be defined in the Nahm dual picture as the integrals over level sets $\{\phi=s_0\}$. That these integrals are independent of $\phi_0$ follows from repeated applications of Stokes theorem.

For the Basu-Harvey equation based on $\CA_4$, one can construct conserved charges $(t(\zeta,\eta),t(\zeta,\eta))$ using the positive definite norm $(\cdot,\cdot)$.

\section{Hermitian bags}\label{sect:HermitianBH}

The Basu-Harvey equation based on the trivial 3-algebra $\FR$ and the 3-algebra $\CA_4$ describe one or two parallel M2-branes ending on M5-branes.  For $n>2$ M2-branes, one needs a generalization of this equation based on hermitian 3-algebras \cite{Terashima:2008sy,Gomis:2008vc,Hanaki:2008cu}.  This equation is a BPS equation of the ABJM model \cite{Aharony:2008ug,Bagger:2008se} just as the Basu-Harvey equation \eqref{eq:BasuFinite} is a BPS equation of the BLG model \cite{Bagger:2007jr,Gustavsson:2007vu}. In this section we will show that our proposed equation \eqref{eq:BHInfinite} for an infinite number of M2-branes arises in the large $n$ limit of this equation.

In order to do this, we first show that the hermitian 3-algebras converge to a sub-algebra of $\CC^\infty(S^3)$ as $n\to\infty$.  The fact that the limit yields a sub-algebra, rather than the whole of $\CC^\infty(S^3)$, places constraints on the bag obtained via the Nahm transform.  We discuss the implications of these constraints at the end of the section: essentially, the bag obtained is invariant under an action of $\sU(1)$, and can be identified with a magnetic bag on $\FR^3$.

\subsection{Equivariant fuzzy 3-sphere}

A hermitian 3-algebra consists of a complex vector space $\CH$ equipped with a 3-bracket $[\,\cdot\,,\,\cdot\,;\,\cdot\,]:\CH^3\to\CH$.  The 3-bracket is anti-symmetric and linear in its first two arguments, and anti-linear in its third argument.  The 3-bracket is required to satisfy the fundamental identity,
\begin{equation}
\label{eq:FIhermitian}
 [[a,b;c],d;e] = [[a,d;e],b;c] + [a,[b,d;e];c] - [a,b;[c,e;d]]
\end{equation}
for all $a,b,c,d,e\in\CH$. The fundamental identity implies that for any $a,b\in\CH$, the maps
\begin{equation}
 D(a,b):c\mapsto [c,a;b]
\end{equation}
span the Lie algebra $\frg_\CH$ of inner derivations of $\CH$.

The conventional ABJM model is built from the hermitian 3-algebra $\sMat_{n\times n}(\FC)$ with bracket 
\begin{equation}
 [C,A;B] = -2n(A\bar BC - C\bar B A) = D(A,B)\acton C~,
\end{equation}
where the bar denotes matrix transposition combined with complex conjugation. Here, we will focus on the sub 3-algebra $\CH_n$ of $(n-1\times n)$-dimensional matrices, which is relevant for the hermitian Basu-Harvey equation.

Note that it is also possible to construct hermitian 3-algebras from 3-Lie algebras.  Given a 3-Lie algebra $\CA$, one defines $\CH=\FC\otimes\CA$ and for the hermitian 3-bracket chooses
\begin{equation}
 [a,b;c] = [a,b,\bar c]~.
\end{equation}
This means for example that the space $\FC\otimes \CC^\infty(S^3)$ of complex functions on $S^3$ forms a hermitian 3-algebra.  We will show below that the large $n$ limit of $\CH_n$ can be identified with a sub-algebra $\CH_\infty$ of $\FC\otimes \CC^\infty(S^3)$.

Consider the following two distinguished elements $W^1,W^2\in\CH_n$:
\begin{equation}
\begin{aligned}
 W^1 &= \frac{1}{\sqrt{n}}\left( \begin{array}{ccccc} 
 0 & \sqrt{1} & 0 & & \vdots \\
 0 & 0 & \sqrt{2} & & \\
 \vdots & & & \ddots & 0 \\
 0 & \cdots & & 0 & \sqrt{n-1}
 \end{array} \right)~, \\
 W^2 &= \frac{1}{\sqrt{n}}\left( \begin{array}{ccccc}
 \sqrt{n-1} & 0 & 0 & & \vdots \\
 0 & \sqrt{n-2} & 0 & & \\
 \vdots & & & \ddots & \\
 0 & \cdots & 0 & \sqrt{1} & 0
 \end{array} \right)~.
\end{aligned}
\end{equation}
These special elements were introduced in \cite{Gomis:2008vc}, where it was noted that they satisfy
\begin{eqnarray}
\label{Ak radius 1}
 \bar W^1 W^1 + \bar W^2 W^2 &=& \frac{n-1}{n} \mathbf{1}_n~, \\
\label{Ak radius 2}
 W^1 \bar W^1 + W^2 \bar W^2 &=& \mathbf{1}_{n-1}~,
\end{eqnarray}
and
\begin{equation}
\label{Ak 3-bracket}
 [W^\alpha,W^\beta;W^\gamma] = 2\eps^{\alpha\beta}\eps^{\gamma\delta}W^\delta~.
\end{equation}
It follows from \eqref{Ak 3-bracket} that the Lie algebra of derivations spanned by $D(-\frac{\mathrm{i}}{2}W^\alpha,W^\beta)$ is $\mathfrak{u}(2)$, and that $W^1,W^2$ transform in the fundamental representation of this Lie algebra.  Thus there is a natural action of $\mathfrak{u}(2)$ on $\CH_n$.

The diagonal sub-algebra $\mathfrak{u}(1)\subset\mathfrak{u}(2)$ is generated by
\begin{equation}
\label{Ak xi}
\Xi = D\left(-\frac{\mathrm{i}}{2} W^\alpha,W^\alpha\right)~,
\end{equation}
and this $\mathfrak{u}(1)$ sub-algebra acts in the following way:
\begin{equation}
\label{eq:Xi action}
\Xi\acton A = \mathrm{i}A
\end{equation}
for all $A\in \CH_n$. The action of the Lie sub-algebra $\mathfrak{su}(2)\subset\mathfrak{u}(2)$ can be summarized by saying that $\CH_n$ transforms in the following representation of $\mathfrak{su}(2)$:
\begin{equation}
\label{Ak functions}
 \CH_n = \overline{\mathbf{n-1}}\otimes\mathbf{n} = \bigoplus_{i=1}^{n-1} \mathbf{2i}~.
\end{equation}

As was noted in \cite{Gomis:2008vc}, the algebraic identities \eqref{Ak radius 1}, \eqref{Ak radius 2} suggest an interpretation of $\CH_n$ as a fuzzy 3-sphere. As we will see now, this interpretation is problematic. It is natural to try to identify the elements $W^1,W^2\in\CH_n$ with the complex functions $w^1=x^1+\mathrm{i} x^2,w^2=x^3+\mathrm{i} x^4$, which satisfy
\begin{equation}
\label{Ainfty radius}
 \bar w^1 w^1 + \bar w^2 w^2 = 1~.
\end{equation}
The hermitian 3-brackets of these functions satisfy the same relations as the 3-brackets of the $W^\alpha$:
\begin{equation}
\label{Ainfty 3-bracket}
 [w^\alpha,w^\beta;w^\gamma]:=\{w^\alpha,w^\beta,\bar{w}^\gamma\}= 2\eps^{\alpha\beta}\eps^{\gamma\delta}w^\delta~,
\end{equation}
where $\{\cdot,\cdot,\cdot\}$ denotes the Nambu-Poisson bracket induced by the canonical volume form on $S^3$. The derivations $D(\frac{\mathrm{i}}{2}w^\alpha,w^\beta)$ therefore span the Lie algebra $\mathfrak{u}(2)$, which can be identified with a Lie sub-algebra of the rotation Lie algebra $\mathfrak{so}(4)\cong \asu(2)\oplus\asu(2)$.

We now explain how this Lie algebra acts on $\FC\otimes \CC^\infty(S^3)$.  We start with the derivation
\begin{equation}
\label{Ainfty xi}
\xi=D\left(-\frac{\mathrm{i}}{2}w^\alpha,w^\alpha\right)=\mathrm{i} \left(w^\alpha\frac{\pa}{\pa w^\alpha}-\bar w^\alpha\frac{\pa}{\pa \bar w^\alpha}\right)~,
\end{equation}
which generates the diagonal $\mathfrak{u}(1)$.  The set of eigenvalues of $\xi$ is $\RZ$, and the identity \eqref{eq:FIhermitian} implies that the eigenspaces of $\xi$ are closed under the hermitian 3-bracket.  In view of \eqref{eq:Xi action} it seems reasonable to identify the large $n$ limit of $\CH_n$ with the hermitian 3-algebra,
\begin{equation}
 \CH_\infty := \{ f:S^3\to\FC\,|\, \xi\acton f = \CL_\xi f = \mathrm{i} f \}~.
\end{equation}
The vector space $\CH_\infty$ may be identified with the space of chiral spinors on the 2-sphere. That is, $\CH_\infty$ is the closure of the span of the polynomials $w^{\alpha_1}\ldots w^{\alpha_{\ell+1}}\bar{w}^{\beta_1}\ldots \bar{w}^{\beta_\ell}$, $\alpha_i,\beta_i=1,2$, $\ell\in\NN$. The space $\CH_\infty$ therefore does {\em not} contain all functions on $S^3$, as would be required by an interpretation of $\CH_n$ as a fuzzy 3-sphere.

Now we consider the action of $\mathfrak{su}(2)$.  It is well-known that $\CH_\infty$ transforms in the following representation of $\mathfrak{su}(2)$:
\begin{equation}
\label{Ainfty functions}
 \CH_\infty = \bigoplus_{i=1}^\infty \mathbf{2i}~.
\end{equation}
This clearly coincides with the $n\to\infty$ limit of \eqref{Ak functions}.  Due to the similarities between equations \eqref{Ak radius 1}, \eqref{Ak radius 2}, \eqref{Ak 3-bracket} and \eqref{Ainfty radius}, \eqref{Ainfty 3-bracket}, it is clear that $\CH_\infty$ is the formal $n\to\infty$ limit of $\CH_n$.

There are obvious parallels to be drawn with the discussion in section \ref{sect:FuzzyFunnel} of the Berezin-Toeplitz quantization of $S^2$: just as $\CCH_n\otimes\overline{\CCH_{n}}$ quantizes the Poisson bracket functions on $S^2$, we have shown that $\CCH_n\otimes\overline{\CCH_{n+1}}$ quantizes the 3-bracket structure on the space of chiral spinors on $S^2$.  It would be interesting to investigate this idea from an analytical point of view, i.e.\ to find analogous formulas to \eqref{eq:approximations} involving Nambu-Poisson and hermitian 3-algebra brackets.

\subsection{The hermitian Basu-Harvey equation}

The BPS equation of the ABJM model, which is conjectured to describe $n$ parallel M2-branes, is the \emph{hermitian Basu-Harvey equation},
\begin{equation}
\label{eq:HBH}
\frac{\dd}{\dd s} Z^\alpha = \frac{\pi^2}{q}[Z^\alpha,Z^\beta;Z^\beta]~,
\end{equation}
where the two functions $Z^1(s),Z^2(s)$ take values in a hermitian 3-algebra. The hermitian 3-algebra chosen in \cite{Gomis:2008vc} was the hermitian 3-algebra of $n\times n$ matrices, however, we saw in the previous section that in order to obtain a reasonable large $n$ limit, it is sensible to restrict attention to the sub-algebra $\CH_n$ of $n-1\times n$ matrices. All irreducible solutions of \eqref{eq:HBH} can be restricted to this sub-algebra \cite{Gomis:2008vc}.

Thus in the large $n$ limit, we obtain $\CH_\infty$-valued functions $z^1(s)$, $z^2(s)$ obeying
\begin{equation}
\label{eq:HBHinf}
\frac{\dd}{\dd s} z^\alpha = \frac{\pi^2}{q}[z^\alpha,z^\beta;z^\beta]~.
\end{equation}
This equation is equivalent to the Basu-Harvey equation \eqref{eq:BHInfinite} if we identify $z^1=t^1+\mathrm{i}t^2$, $z^2=t^3+\mathrm{i} t^4$.   The natural range for the variable $s$ is here $[0,v)$, and the boundary \eqref{nahmboundary} can be rewritten as
\begin{equation}
\label{eq:HBHBC}
z^{\alpha}=\frac{ w^\alpha}{2\pi}\sqrt{\frac{q}{v-s}}+\CO((v-s)^{\frac{1}{2}}) \mbox{ as } s\to v ~.
\end{equation}
A self-dual string bag on $\FR^4$ can be obtained by applying the Nahm transform to any solution of \eqref{eq:HBHinf}, \eqref{eq:HBHBC} as in subsection \ref{sect:SDStransform}.

However, the fact that $z^1,z^2$ take values in $\CH_\infty$ and not the full function space $\CC^\infty(S^3)$ imposes constraints on the bag obtained.  We will now show that bags resulting from solutions to \eqref{eq:HBHinf}, \eqref{eq:HBHBC} are invariant under a certain $\sU(1)$-action, and moreover that they are equivalent to magnetic bags on $\FR^3$.

Let $\eta$ be the following vector field on $\FR^4$:
\begin{equation}
\eta = y^1\frac{\pa}{\pa y^2} - y^2\frac{\pa}{\pa y^1} + y^3\frac{\pa}{\pa y^4} - y^4\frac{\pa}{\pa y^3}~.
\end{equation}
The fact that $\CL_\xi z^\alpha=\mathrm{i}z^\alpha$ implies that
\begin{equation}
 \CL_\xi t^1=-t^2~,\quad
 \CL_\xi t^2=t^1~,\quad 
 \CL_\xi t^3=-t^4~,\quad
 \CL_\xi t^4=t^3~.
\end{equation}
It follows that the push-forward of $\xi$ under the map $t:S^3\times[0,v)\to\Omega\subset\FR^4$ is $\eta$: $t_\ast\xi=\eta$.  Now the coordinate function $s$ and the 3-form $\omega$ on $S^3\times[0,v)$ satisfy $\CL_\xi s=0$ and $\CL_\xi \omega=0$; therefore the function $\phi$ and 3-form $h$ obtained under the Nahm transform satisfy $\CL_\eta\phi=0$, $\CL_\eta h=0$.  This means that $\phi$ and $h$ are invariant under the action of $\sU(1)$ generated by $\eta$, and similarly the bag surface $\Sigma=\dpar \Omega$ is $\sU(1)$-invariant.

\subsection{Magnetic bags from self-dual string bags}

Since the bag on $\FR^4$ obtained from a solution to \eqref{eq:HBHinf}, \eqref{eq:HBHBC} is $\sU(1)$-invariant, it is natural to try to identify it with some configuration on the quotient space.  It is well-known that $\FR^4/\sU(1)\cong\FR^3$; standard coordinates on $\FR^3$ are defined by the $\sU(1)$-invariant functions,
\begin{equation}
\label{R4 to R3}
r^i := \left( \begin{array}{cc} y^1-\mathrm{i}y^2 & y^3-\mathrm{i} y^4 \end{array} \right) \sigma^i 
\left( \begin{array}{cc} y^1+\mathrm{i}y^2 \\ y^3+\mathrm{i} y^4 \end{array} \right)~.
\end{equation}
We will denote this projection from $\FR^4$ to $\FR^3$ by $\pi$. When restricted to $S^3\embd \FR^4$, the projection $\pi$ is nothing but the Hopf fibration $S^1\rightarrow S^3\stackrel{\pi}{\rightarrow} S^2$. 

Let us return to the solution $(h,\phi)$ constructed in the previous subsection. The function $\phi$ is $\sU(1)$-invariant, so it must be the pull-back of some function $\psi$ on (a subset of) $\FR^3$.  The 3-form $h$ cannot be the pull-back of a 3-form on $\FR^3$, because $\iota_\eta h\neq 0$.  However, the 2-form $\iota_\eta h$ satisfies $\iota_\eta(\iota_\eta h)=0$ and $\CL_\eta(\iota_\eta h)=0$, so it is the pull-back of some 2-form $f$ on $\FR^3$.  This 2-form $f$ is closed, because
\begin{equation}
\pi^\ast\dd f = \dd \pi^\ast f = \dd \iota_\eta h = \CL_\eta h + \iota_\eta\dd h = 0~.
\end{equation}

Now we will determine what equation $(f,\psi)$ must satisfy.  It can be shown that, for any 1-form $u$ on $\FR^3$,
\begin{equation}
\ast_4 \pi^* u = \theta\wedge\pi^\ast(\ast_3 u)~,
\end{equation}
where $\pi:\FR^4\to\FR^3$ is the projection, $\ast_4$ and $\ast_3$ are the Hodge star operators on $\FR^4$ and $\FR^3$ with respect to the standard flat metrics, and
\begin{equation}
\theta := \frac{1}{y^\mu y^\mu}\left( -y^2\dd y^1 + y^1\dd y^2 - y^4\dd y^3 + y^3\dd y^4 \right)~.
\end{equation}
Since $\iota_\eta \theta = 1$, it follows that
\begin{equation}
\pi^\ast f = \iota_\eta h = \iota_\eta (\ast_4\dd\phi) = \iota_\eta (\ast_4 \pi^\ast \dd\psi) = \iota_\eta(\theta\wedge \pi^\ast (\ast_3\dd\psi)) = \pi^\ast (\ast_3\dd\psi)~,
\end{equation}
Therefore $(f,\psi)$ satisfy $f=\ast_3\dd\psi$ and define a magnetic bag on $\FR^3$.

Conversely, given any magnetic bag $(f,\psi)$ on $\FR^3$, a self-dual string bag on $\FR^4$ can be obtained by setting $\phi=\pi^\ast\psi$, $h=\theta\wedge\pi^\ast f$.  One can check that $(h,\phi)$ satisfy the self-dual string equation:
\begin{equation}
\ast_4\dd\phi = \ast_4\pi^\ast\dd\psi = \theta\wedge\pi^\ast(\ast_3\dd\psi) = \theta\wedge\pi^\ast f = h~.
\end{equation}
Moreover, $h$ is closed, because
\begin{equation}
\iota_\eta \dd h = \iota_\eta(\dd\theta\wedge\pi^\ast f -\theta\wedge\dd\pi^\ast f) = \iota_\eta(\dd\theta\wedge\pi^\ast f) = 0~,
\end{equation}
where in the last equality we have used the facts that $\iota_\eta\pi^\ast f=0$ and $\iota_\eta \dd\theta=0$.  Any 4-form whose inner derivative with $\eta$ vanishes must be zero, so it must be the case that $\dd h=0$.  Thus we have established a bijective correspondence between $\sU(1)$-invariant self-dual string bags and magnetic bags.

This correspondence can also be seen at the level of $\Pi_\omega$-Nahm data.  Viewed as functions on $S^3\times [0,v)$, $t^i=\bar{z}^\alpha\sigma^i_{\alpha\beta} z^\beta$ are invariant under the group $\sU(1)$ generated by $\xi$. The space of $\sU(1)$-invariant functions on $S^3$ can be identified with the space of functions on $S^2$, via the Hopf fibration.  This function space is equipped with a Poisson bracket, defined via
\begin{equation}
4\{f,g\}_{S^3}~ \iota_\xi\omega = \dd f\wedge\dd g,
\end{equation}
where $f,g$ are any $\sU(1)$-invariant functions on $S^3$.  The Poisson bracket can be lifted to $S^3\times[0,v)$ by wedging both sides of this equation with $\dd s$.  It is straightforward (but tedious) to check that the hermitian Basu-Harvey equation \eqref{eq:HBHinf} implies that $t^i$ satisfy the Nahm equation,
\begin{equation}
\frac{\dd }{\dd s} t^i = \frac{4\pi^2}{q}\frac{1}{2}\eps_{ijk}\{t^j,t^k\}_{S^3}~,
\end{equation}
Altogether, we have proved the following theorem:

\begin{thm}
Up to gauge equivalence, we have one-to-one correspondences between the following sets:
\begin{equation}
\label{commutative diagram}
\begin{array}{ccc}
\mbox{$\CH_\infty$ hermitian Basu-Harvey data} & \longleftrightarrow & \mbox{$\sU(1)$-invariant bags on }\FR^4 \\
\updownarrow & & \updownarrow \\
\mbox{$\Pi_\omega$-Nahm data for magnetic bags} & \longleftrightarrow & \mbox{magnetic bags on }\FR^3
\end{array}
\end{equation}
\end{thm}

\section{Loop space self-dual string bags}\label{sect:LSSDS}

Recent investigations of self-dual strings have made use of loop space, cf.\ \cite{Gustavsson:2008dy,Saemann:2010cp,Palmer:2011vx}.  We will show in this section that the Nahm transform for self-dual string bags has a formulation in loop space; this sets the transform in a wider context.  This formulation makes essential use of naturally defined Poisson-like brackets on 1-forms and loop space, so we begin by reviewing these constructions.

\subsection{Poisson-like structures on 1-forms}\label{subsec:PoissonForms}

To any 1-form $\alpha$ on $S^3$, a vector field $X_\alpha$ can be associated via the equation
\begin{equation}
 \dd \alpha = \iota_{X_\alpha}\omega~.
\end{equation}
It follows directly that $\CL_{X_\alpha}\omega=0$, so the vector field $X_\alpha$ is volume-preserving or divergence-free. This generalizes the relationship between functions and vector fields on a symplectic manifold.  The 1-form $\alpha$ is called a Hamiltonian 1-form and $X_\alpha$ is the corresponding Hamiltonian vector field, while the volume form $\omega$ is sometimes called a 2-plectic form.\footnote{More generally, a closed non-degenerate $p+1$-form $\omega$ on a manifold is called a $p$-plectic form, and one can speak of Hamiltonian $p-1$-forms and vector fields.  It is not true in general that every $p-1$-form is Hamiltonian, however, on $S^3$ every 1-form is Hamiltonian.}

There are two obvious generalizations of the Poisson bracket on 1-forms \cite{Baez:2008bu}: the {\em hemi-bracket} is defined as
\begin{equation}
 \{\alpha,\beta\}_h:=\CL_{X_\alpha}\beta~,
\end{equation}
and the {\em semi-bracket} is given by
\begin{equation}
  \{\alpha,\beta\}_s:=\iota_{X_\alpha}\iota_{X_\beta}\omega~.
\end{equation}
The hemi-bracket satisfies the Jacobi-identity but it is not antisymmetric, while the semi-bracket is anti-symmetric but does not satisfy the Jacobi-identity.\footnote{Note that for so-called exact multisymplectic manifolds, which $S^3$ is not, a further bracket can be constructed that is both antisymmetric and satisfies the Jacobi identity \cite{Forger:2002aa}.}  The difference between the hemi- and semi-brackets is an exact 1-form: 
\begin{equation}\label{eq:diffHemiSemi}
 \{\alpha,\beta\}_h-\{\alpha,\beta\}_s=\dd \iota_{X_\alpha} \beta~.
\end{equation}
It follows that $\{\alpha,\beta\}_h$ and $\{\alpha,\beta\}_s$ induce the same vector field on $S^3$.  In fact, one has that
\begin{equation}
 X_{\{\alpha,\beta\}_h} = X_{\{\alpha,\beta\}_s} = [X_\alpha,X_\beta]~.
\end{equation}

On $S^3$, we may write $\alpha=\dd\theta^i\,\alpha_i$ and $\beta=\dd\theta^i\,\beta_i$, where $\theta^i$, $i=1,2,3$, denote again the canonical angles. Then the semi-bracket explicitly reads as
\begin{equation}
\{\alpha,\beta\}_s=\iota_{X_\alpha}\iota_{X_\beta}\omega=\dd\theta^i~\frac{\eps^{jkl}}{\sin^2\theta^1 \sin\theta^2}~\der{\theta^{[j}}\alpha_{i]}~\der{\theta^k} \beta_l~.
\end{equation}

\subsection{Poisson structures on loop space}

Consider now the free loop space $\CL S^3$ of $S^3$, whose elements are given by loops $\theta:S^1\to S^3$. The tangent space at a loop $\theta$ is given by
\begin{equation}
 T_\theta\CL S^3=\CC^\infty(S^1,\theta^* TS^3)~.
\end{equation}
Thus, we will write tangent vectors as 
\begin{equation}
 \xi=\oint \dd \tau~ \xi^i(\theta,\tau)\,\delder{\theta^i(\tau)}=\oint \dd \tau~ \xi^{i\tau}(\theta)\,\delder{\theta^{i\tau}}~,
\end{equation}
and dual 1-forms as 
\begin{equation}
 \chi=\oint \dd \tau~ \chi_{i\tau}(x)\,\delta\theta^{i\tau}~,
\end{equation}
with $\langle \delta \theta^{i\tau},\delder{\theta^{j\sigma}}\rangle=\delta^i_j\delta(\tau-\sigma)$.  The total differential is
\begin{equation}
 \delta=\oint \dd \tau~\delta \theta^{i\tau}\delder{\theta^{i\tau}}~.
\end{equation}

Reparameterizations of a loop $\theta(\tau)$ are generated by the vector fields
\begin{equation}
\label{eq:trivial tangent vector}
\Gamma= \oint \dd \tau~ \gamma(\tau)\,\dot\theta^{i}(\tau)\,\delder{\theta^{i\tau}}~,
\end{equation}
where $\gamma$ is a function of $\tau$, transforming appropriately under reparameterizations. The quotient of the free loop space by this action is the space of unparameterized loops\footnote{Strictly speaking, we restrict ourselves to the space of singular knots, see \cite{0817647309} for details.}, which we denote by $\CL S^3$. We will still describe these loops by maps $\theta:S^1\rightarrow S^3$, but we will ensure that all our formulas are reparameterization invariant. Moreover, we impose the relations
\begin{equation}
 \dot{\theta}^{i}(\tau)\delder{\theta^{i\tau}}=\dot{\theta}_{i}(\tau)\delta\theta^{i\tau}=0\quad\forall\tau\in S^1~.
\end{equation}

The \emph{transgression map} \cite{0817647309} sends $p$-forms on a manifold $M$ to $p-1$-forms on its loop space $\CL M$: One of the $p$-form's indices can be contracted with the tangent vector to the loop under consideration. For a $p$-form $\omega=\frac{1}{p!}\omega_{i_1\cdots i_p}(\theta)\dd \theta^{i_1}\wedge\cdots\wedge \dd \theta^{i_p}$ we have explicitly the following local expression:
\begin{equation}\label{eq:Transgression}
 (\CT \omega)(\theta)=\oint_x \dd \tau~ \tfrac{1}{(p-1)!}\,\omega(\theta(\tau))_{i_1\cdots i_{p}}~\dot{\theta}^{i_p}~\delta \theta^{i_1\tau}\wedge \cdots \wedge \delta \theta^{i_{p-1}\tau}~.
\end{equation}
Note that $\CT\omega$ is reparameterization invariant. Furthermore, the transgression map is a chain map, which means that closed forms are mapped to closed forms and exact forms are mapped to exact forms.  In particular, the transgression of an exact 1-form is zero: 
\begin{equation}
 \CT(\dd f)=\oint \dd \tau~ \dot\theta^i(\tau)\left.\pa_if\right|_{\theta(\tau)}=\oint \dd \tau~ \dder{\tau} f(\theta(\tau))=0~.
\end{equation}
Note that the transgression map is not surjective. We will call forms on $\CL S^3$ which are in the image of $\CT$ {\em ultralocal}. Moreover, we will call forms on $\CL S^3$ that can be written in terms of a single loop integral {\em local}.

Consider now the standard volume form $\omega$ on $S^3$. The 2-form $\CT\omega$ is closed and non-degenerate, and therefore the volume form (or 2-plectic structure) on $S^3$ is lifted by the transgression map to a symplectic structure on $\CL S^3$ \cite{0817647309}.  Thus to any function $f$ on loop space, one can associate a Hamiltonian vector field $X_f$ in the usual way, and a Poisson bracket can be defined on loop space by the formula $\{f,g\}=\iota_{X_f}\iota_{X_g}\CT\omega$.

Interestingly, the components of the Hamiltonian vector field of a 1-form $\alpha\in \Omega^1(S^3)$ with respect to a 2-plectic form $\omega$ are identical to those of the Hamiltonian vector field of $\CT\alpha$ with respect to $\CT \omega$:
\begin{equation}
 X_\alpha= X^i_\alpha\der{\theta^i}~~~\Rightarrow~~~X_{\CT\alpha}=\oint \dd \tau~ X^i_{\alpha}(\theta(\tau))\,\delder{\theta^i(\tau)}~.
\end{equation}
This implies that the transgression maps both semi- and hemi-brackets on $(M,\omega)$ to the Poisson bracket on $(\CL S^3,\CT\omega)$:
\begin{equation}
 \CT\{\alpha,\beta\}_h=\CT\{\alpha,\beta\}_s=\{\CT\alpha,\CT\beta\}_{\CT\omega}~.
\end{equation}
Note that the transgressions of the hemi- and semi-brackets agree because their difference is an exact 1-form, and the transgression of exact 1-forms is zero.

\subsection{The Basu-Harvey equation in loop space}

We have now all the preliminaries covered to discuss the Basu-Harvey equation on loop space.  If $t^\mu(s)$ solve the Basu-Harvey equation \eqref{eq:BHInfinite} then the vector fields $D(t^\mu,t^\nu)$ solve
\begin{equation}
\label{eq:VFinfBH}
\eps_{\mu\nu\kappa\lambda}\frac{\dd}{\dd s}D(t^\kappa,t^\lambda)=\frac{4\pi^2}{ q}~[D(t^{\mu},t^{\kappa}),D(t^{\nu},t^{\kappa})]~.
\end{equation}
Consider the following 1-forms on $S^3$:
\begin{equation}
 t^{\mu\nu}:=t^{[\mu}\dd_{S^3} t^{\nu]}~,
\end{equation}
where $\dd_{S^3}$ denotes the exterior derivative on $S^3$, i.e.\ $\dd_{S^3} t^\mu(\theta^i,s)=\partial_i t^\mu\,\dd \theta^i$. The vector fields $D(t^\mu, t^\nu)$ are Hamiltonian vector fields associated to these 1-forms,
\begin{equation}
 X_{t^{\mu\nu}}=D(t^{\mu},t^{\nu})~.
\end{equation}
Since two Hamiltonian one-forms yielding the same Hamiltonian vector field on $S^3$ can differ only by an exact form $\gamma$, the Basu-Harvey equation implies the following equation for the 1-forms $t^{\mu\nu}$:
\begin{equation}
\eps_{\mu\nu\rho\lambda}\frac{\dd}{\dd s}t^{\rho\lambda}=\frac{4\pi^2}{ q}~\{t^{\mu\kappa},t^{\nu\kappa}\}_s+\gamma~,
\end{equation}
where $\dd \gamma=0$. Let us now switch to loop space via the transgression map \eqref{eq:Transgression}. The 1-forms $t^{\mu\nu}$ are mapped to the following functions on $\CL S^3\times[0,v)$:
\begin{equation}
 \trng^{\mu\nu}(\theta,s):=\CT t^{\mu\nu}=\oint\dd\tau\, t^\mu(\theta^a(\tau),s)\frac{\dd}{\dd\tau}t^\nu(\theta^a(\tau),s)~.
\end{equation}
These functions satisfy the loop space Basu-Harvey equation,
\begin{equation}\label{eq:LSinfBH}
\eps_{\mu\nu\kappa\lambda}\frac{\dd}{\dd s}\trng^{\kappa\lambda}=\frac{4\pi^2}{ q}~\{\trng^{\mu\kappa},\trng^{\nu\kappa}\}_{\CT\omega}~,
\end{equation}
where $\{\cdot,\cdot\}_{\CT\omega}$ denotes the natural Poisson structure on $\CL S^3$ induced by the transgressed volume form on $S^3$.

We would like to point out that while the Basu-Harvey equation implies equation \eqref{eq:VFinfBH}, the converse is not true. For example, the solution $t^1=\frac{e}{s}$, $t^2=t^3=t^4=0$, $e\in \CA$, to equation \eqref{eq:VFinfBH} does not satisfy the Basu-Harvey equation. Furthermore, one can exploit Gustavsson's observation \cite{Gustavsson:2008dy} that \eqref{eq:VFinfBH} is equivalent to two copies of the Nahm equation; these are obtained by projecting out the self-dual and anti-self-dual parts of the antisymmetric tensors $D(t^\kappa,t^\lambda)$ using the 't Hooft tensors.  Ashtekar et al.\ have shown \cite{Ashtekar:1987qx} that self-dual and anti-self-dual Einstein metrics can be constructed from any solution of the Nahm equation based on a Lie algebra of volume-preserving diffeomorphisms.  This means in particular that there are two Einstein metrics naturally associated to solutions of \eqref{eq:VFinfBH}, one self-dual and the other anti-self-dual.  A short calculation shows that these metrics both 
coincide with the flat metric on $\FR^4$ for all solutions of \eqref{eq:VFinfBH} obtained from the Basu-Harvey equation \eqref{eq:BHInfinite}.  Thus the space of solutions of the Basu-Harvey equation forms a special subspace of the space of solutions of \eqref{eq:VFinfBH}.

\subsection{Constructing loop space self-dual strings}

Let us now come to the loop space version of the self-dual string equation. Recall that the diffeomorphism $t:S^3\times \CI\rightarrow \Omega$ induces a foliation of $\Omega$ with leaves $\Sigma_{\phi}\cong S^3$, and for the interval $\CI=[0,v)$, $v\in\FR^+$, this foliation can be considered to be a fibration $\Omega\rightarrow\FR^+$ with fiber $\Sigma_{\phi(r)}$. Replacing $S^3$ with its loop space, we have the following induced map
\begin{equation}
 \trng:\CL S^3\times \CI\rightarrow \CL\Omega~,~~~\trng(\theta,s,\tau):=t(\theta(\tau),s)~,
\end{equation}
where $\theta$, $s$ and $\tau$ are coordinates on $S^3$, $\CI$ and $S^1$, respectively. This map is a diffeomorphism only between $\CL S^3\times \CI$ and its image in $\CL\Omega$. The latter space consists of loops that lie entirely in the fibers of $\Omega\rightarrow \FR^+$, and we effectively have a diffeomorphism $\trng: \CL S^3\times \CI\rightarrow \CL \Sigma\times \FR^+$ together with its inverse $\urng$. 

As before, we would like to construct a field strength $\CF$ together with a Higgs field $\varPhi$ by pulling back the transgressed volume form $\CT\omega$ and the coordinate function $s$ along $\urng$. The pull-back 2-form $\CF:=\frac{q}{2\pi^2}\urng^* \CT \omega$ has components 
\begin{equation}
\CF=\tfrac{1}{2}\oint \dd\tau~\CF_{(\mu\tau)(\nu\tau)}\delta y^{\mu\tau}\wedge\delta y^{\nu\tau}~.
\end{equation}
Note that the transgression $\CT \omega$ is by definition ultralocal, and so is the pull-back $\CF=\urng^*\CT\omega$ along inverses $\urng$ of induced maps $\trng$.

A full loop space self-dual string equation has been derived in \cite{Palmer:2011vx}. For an abelian local field strength, it reduces to the following equation:
\begin{equation}\label{eq:exLoopSpaceSDS}
 \CF_{(\mu\tau)(\nu\tau)}=\eps_{\mu\nu\kappa\lambda}~\frac{\yd^{\kappa\tau}}{|\yd^\tau|}\delder{y^{\lambda\tau}}\varPhi~,
\end{equation}
where $\varPhi$ is a $\au(1)$-valued function on $\CL\Omega$. Note that in the case of the non-abelian version of this equation, a formulation in terms of local forms is no longer gauge invariant and therefore not very useful. In the abelian case, however, gauge transformations act trivially on $\CF$ and $\varPhi$.

The right-hand side of \eqref{eq:exLoopSpaceSDS} can be understood as a Hodge-star for certain local forms generalized to loop space\footnote{Recall that there is no Hodge-star operation on general forms, because loop space is infinite dimensional.}. We define on $\CL\Omega\subset \CL\FR^4$:
\begin{equation}
\begin{aligned}
 * \oint \dd \tau~\alpha_{\mu_1\ldots \mu_p,\tau}~\delta y^{\mu_1\tau}&\wedge \cdots \wedge \delta y^{\mu_p\tau}:=\\
&\oint \dd \tau~\frac{(-1)^{p+1}}{p!}\alpha_{\mu_1\ldots \mu_p,\tau}\eps^{\mu_1\ldots \mu_{4}}\frac{\yd_{\mu_{p+1}\tau}}{|\yd^\tau|}\delta y^{\mu_{p+2}\tau}\wedge \cdots\wedge \delta y^{\mu_4\tau}~,
\end{aligned}
\end{equation}
where $0\leq p\leq 3$. One easily verifies that $*^2=\id$. The loop space self-dual string equation \eqref{eq:exLoopSpaceSDS} then reduces to $\CF=*\delta \varPhi$. 

We now restrict to $\CL\Sigma\times \FR^+$, which is diffeomorphic to $\CL S^3\times \CI$. As before, the $\Sigma_\varPhi$ are the level sets of the Higgs field $\varPhi$, and $\varPhi(\theta,r)=\varPhi(r)$. The total differential $\delta$ on $\CL\Sigma\times \FR^+$ reduces such that the equation $\CF=*\delta \varPhi$ becomes
\begin{equation}\label{eq:6.27}
\CF=\oint \dd\sigma~\eps_{\mu\nu\kappa\lambda}\yd^{\kappa\sigma}\left. \frac{\partial \varPhi(x)}{\partial x^\lambda}\right|_{x=y(\sigma)}\delta y^{\mu\sigma}\wedge\delta y^{\nu\sigma}~.
\end{equation}
Let us now verify that the fields $\CF=\frac{q}{2\pi^2}\urng^*\CT \omega$ and $\varPhi=s\circ \urng$ indeed solve this equation, where $u_\circ$ is the inverse of the diffeomorphism $t_\circ$.  Equation \eqref{eq:6.27} is equivalent to the following equation on $\CL S^3\times\CI$:
\begin{multline}
\frac{q}{2\pi^2}\CT \omega = \oint \dd\sigma~\eps_{\mu\nu\kappa\lambda}\trngd^{\kappa}(\sigma)\left. \frac{\partial s}{\partial x^\lambda}\right|_{x=t(\theta(\sigma),s)} \times \\
\left[ \frac{1}{2}\oint \dd \tau'\oint \dd \tau'' \frac{\delta \trng^{\mu\sigma}}{\delta \theta^{i\tau'}}\frac{\delta \trng^{\nu\sigma}}{\delta\theta^{j\tau''}} \delta\theta^{i\tau'}\wedge\delta \theta^{j\tau''} +\oint \dd \tau \frac{\partial \trng^{\mu\sigma}}{\partial s} \frac{\delta \trng^{\nu\sigma}}{\delta \theta^{i\tau}} \dd s\wedge\delta\theta^{i\tau}\right]~.
\end{multline}
The right hand side of this equation can be simplified using the identities,
\begin{equation}
 \frac{\delta \trng^{\mu\sigma}}{\delta \theta^{i\tau}}= \left. \frac{\partial t^{\mu}(\chi,s)}{\partial \chi^{i}}\right|_{\chi=\theta(\sigma)}\delta(\sigma-\tau)~~,~~~\trngd^{\mu}(\sigma)=\left. \frac{\partial t^{\mu}(\chi,s)}{\partial \chi^{i}}\right|_{\chi=\theta(\sigma)}\dot \theta^{i}(\sigma)~,
\end{equation}
yielding
\begin{equation}
\oint \dd\sigma\,\eps_{\mu\nu\kappa\lambda}\, \partial_\lambda s\, \dot\theta^k(\sigma)\,\partial_k t^\kappa
\left[ \frac{1}{2} \partial_i t^\mu \partial_j t^\nu \delta\theta^{i\sigma}\wedge\delta \theta^{j\sigma} + \partial_s t^{\mu} \partial_i t^\nu \dd s\wedge\delta\theta^{i\sigma}\right].
\end{equation}
On substituting for $\pa_s t^\mu$ using the Basu-Harvey equation \eqref{eq:BHInfinite}, the second term in the square bracket becomes an expression which vanishes due to $\frac{\partial s}{\partial t^\mu}\frac{\partial t^\mu}{\partial \theta^a}=0$.  The first term in the square bracket can be rearranged using the Basu-Harvey equation \eqref{eq:BHInfinite} to give
\begin{equation}
\frac{q}{2\pi^2} \oint\dd\tau \frac{1}{2}\sin^2\theta^1(\tau)\sin\theta^2(\tau)\eps_{ijk} \dot\theta^i(\tau)\delta\theta^{j\tau}\delta\theta^{k\tau}.
\end{equation}
This expression is clearly equal to $\frac{q}{2\pi^2}\CT \omega$, so the Basu-Harvey equation \eqref{eq:BHInfinite} implies the loop space self-dual string equation \eqref{eq:6.27}.

\section{Magnetic domains in higher dimensions}\label{sec:higher}

Although string- and M-theory only motivate the study of magnetic domains in three and four dimensions, it is still interesting to consider higher-dimensional generalizations of these objects. 

\subsection{From higher BPS equations to magnetic domains}

Recall that the curvature 2-form $f$ of a magnetic domain $\Omega$ in three dimensions defines topologically a vector bundle over $\Omega$. Using repeatedly the Poincar\'e lemma, we obtain gauge potentials on patches of a covering of $\Omega$ and transition functions on overlaps of patches. Analogously, the curvature 3-form $h$ of a magnetic domain in four dimensions defines a gerbe, with 2-form potentials on patches etc.\ Vector bundles and gerbes are examples of so-called {\em $k$-gerbes} with $k=0$ and $k=1$, respectively. In general, $k$-gerbes are defined in terms of curvature $k+2$-forms, with associated $k+1$-form potential etc.

Let us now generalize our previous discussion to magnetic domains in $k+3$-dimensions, which are described in terms of an abelian Higgs field together with the curvature $k+2$-form $g$ of a $k$-gerbe. As before, we can obtain a Bogomolny bound from the energy functional 
\begin{equation}
E=\tfrac{1}{2}\int_\Omega~ g\wedge * g+\dd\phi\wedge *\dd\phi~,
\end{equation}
where $\Omega\subset\FR^{k+3}$ and $g=\dd c$ for some $k+1$-form potential $c$. The Bogomolny bound then becomes
\begin{equation}
E=\int_\Omega~ \tfrac{1}{2}|\dd\phi-*g|^2+\dd\phi\wedge g\ge vq~,
\end{equation}
where $q:=\int_{S^{k+2}_\infty}g$, and the bound is saturated if 
\begin{equation}
g=*\dd\phi~.
\end{equation}

On a $k+2$-dimensional orientable manifold $M$, a volume form $\omega$ yields a $k+1$-plectic structure, i.e.\ a non-degenerate and closed $k+2$-form. This form can be inverted to a multivector field, which defines a Nambu-Poisson structure on $\CC^\infty(M)$. That is, we have a $k+2$-ary bracket $\{\cdot,\,\cdots,\cdot\}$ which is linear in each argument, totally antisymmetric and satisfies the obvious generalizations of the fundamental identity \eqref{eq:FI3Alg} and the Leibniz rule \eqref{eq:Leibniz}. 

The $\Pi_\omega$-Nahm data is here given by a $k+3$-tuple of functions $t^m$, $m=1,\ldots,k+3$ on a $k+1$-plectic manifold $M$, which solve the higher Nahm equation
 \begin{equation}\label{eq:higherNahm}
\frac{\dd t^{m_1}}{\dd s}=\frac{{\rm vol}(S^{k+2})}{(k+2)! q}~\eps_{m_1\dots m_{k+3}}\{t^{m_2},\dots,t^{m_{k+3}}\}~,
\end{equation}
where the volume of the unit $S^{k+2}$-sphere is
\begin{equation}
{\rm vol}(S^{k+2})=\frac{2\pi^\frac{k+3}{2}}{\Gamma(\frac{k+3}{2})}~~.
\end{equation}
The generalization of theorems \ref{thm:2} and \ref{thm:4} now reads as
\begin{thm}
\label{thm:6}
Up to gauge equivalence, there is a one-to-one correspondence between
\vspace{-0.2cm}
\begin{itemize}
 \setlength{\itemsep}{-1mm}
 \item sets of $\Pi_\omega$-Nahm data with the property that the map from $M\times \CI$ to $\Omega\subset\FR^{k+3}$ defined by the $t^m$ is a diffeomorphism $t:M\times \CI\rightarrow \Omega$, and
 \item magnetic domains $\Omega$ that are diffeomorphic to $M\times \CI$, where the restriction of the $k+2$-form curvature $g$ to any slice $M\times\{s_0\}$ has the same volume type as $\omega$ and $\CI$ is the range of $\phi$. Explicitly, there is a diffeomorphism $u:\Omega\rightarrow M\times \CI$ such that $g=\frac{q}{{\rm vol}(S^{k+2})}u^*\omega$ and $\phi=s\circ u$ on $\Omega$. 
\end{itemize}
\end{thm}

The proof is an obvious generalization of the proofs of theorems \ref{thm:2} and \ref{thm:4}. 

\

For a $k+2$-dimensional magnetic bag in $\FR^{k+3}$, the boundary condition for the Higgs field reads as
\begin{equation}
\phi=v-\frac{q}{{\rm vol}(S^{k+2}) r^{k-1}(k-1)}+\CO(r^{-k})
\end{equation}
for $r\rightarrow \infty$. The corresponding boundary condition for the $\Pi_\omega$-Nahm data is
 \begin{equation}
t^{i}= x^i\left(\frac{q}{{\rm vol}(S^{k+2})(k-1)(v-s)}\right)^{\frac{1}{k-1}}+\CO(s^{\frac{k}{1-k}})
\end{equation}
as $s\rightarrow v$. Instead of discussing examples of such magnetic domains in more detail, let us comment on the relation of our equations to $L_\infty$-algebras.

\subsection{Comments on the relation to strong homotopy Lie algebras}\label{ssec:SHalgebras}

The appearances of higher brackets in our equations suggests to look for a relationship to strong homotopy Lie algebras or $L_\infty$-algebras for short. Roughly speaking, an $L_\infty$-algebra is a graded vector space together with brackets with arbitrarily many arguments that satisfy homotopy Jacobi identities. They are the most natural generalization of (differential) Lie algebras to vector spaces endowed with higher brackets. Let us briefly recall the exact definition \cite{Lada:1992wc,Lada:1994mn}.

For this, we need two notions, both related to permutations $\sigma\in S_n$, where $S_n$ is the symmetric group\footnote{i.e.\ the permutation group} of $n$ elements. First, a permutation $\sigma\in S_{i+j}$ is called an {\em $(i,j)$-unshuffle}, if the first $i$ and the last $j$ images of $\sigma$ are ordered: $\sigma(1)<\cdots<\sigma(i)$ and $\sigma(i+1)<\cdots<\sigma(i+j)$. Second, given a graded vector space $L$, one defines the {\em Koszul sign} $\eps(\sigma;x_1,\cdots,x_n)$, $x_i\in L$ via the equation
\begin{equation}
 x_1\wedge \cdots \wedge x_n=\eps(\sigma;x_1,\cdots,x_n)\,x_{\sigma(1)}\wedge \cdots \wedge x_{\sigma(n)}~,
\end{equation}
in the free graded algebra $\wedge (x_1,\cdots,x_n)$.

An {\em $L_\infty$-algebra}, or {\em strong homotopy Lie algebra}, is a graded vector space $L$ endowed with totally antisymmetric maps (or products or brackets) $\mu_k:L^{\wedge k}\rightarrow L$ for $k\in \NN^*$. These maps are of grading $2-k$ and satisfy the homotopy Jacobi identities
\begin{equation}\label{eq:homotopyJacobi}
 \sum_{i+j=n}\sum_\sigma {\rm sgn}(\sigma)\eps(\sigma;x_1,\cdots,x_n)(-1)^{i\cdot j}\mu_{j+1}(\mu_i(x_{\sigma(1)},\cdots,x_{\sigma(i)}),x_{\sigma(i+1)},\cdots,x_{\sigma(i+j)})=0
\end{equation}
for $m\in \NN^*$, where the sum over $\sigma$ is taken over all $(i,j)$ unshuffles. If the graded vector space underlying an $L_\infty$-algebra $L$ is concentrated in degrees $k=0,\ldots,n-1$, i.e.\ $L=\oplus_{k\in\RZ} L_k$ with $L_k=0$ unless $0\leq k\leq n-1$, then $L$ is a {\em Lie $n$-algebra}. In a Lie $n$-algebra, we have $\mu_i=0$ for $i>n+1$. Note that the homotopy Jacobi identity implies $\mu_1^2=0$, such that $\mu_1$ is a differential. Therefore, a Lie $1$-algebra is an ordinary Lie algebra. A Lie 2-algebra, or 2-term $L_\infty$-algebra, consists of two vector spaces $V_0$ and $V_1$ with differential $\mu_1:V_1\rightarrow V_0$, a binary map $\mu_2:V_i\times V_j\rightarrow V_{i+j}$, $i,j,i+j=0,1$ and a ternary map $\mu_3:V_0\times V_0\times V_0\rightarrow V_1$, all satisfying \eqref{eq:homotopyJacobi}.

Strong homotopy Lie algebras appear in modern deformation theory. Here, the definition of the deformation functor involves the so-called {\em homotopy Maurer-Cartan equations} \cite{Merkulov:1999aa,Lazaroiu:2001nm} on an element $\phi$ of an $L_\infty$-algebra $L$:
\begin{equation}\label{eq:MCeqs}
 \sum_{i=1}^\infty \frac{(-1)^{i(i+1)/2}}{i!}\mu_i(\phi,\cdots,\phi)=0~.
\end{equation}
These equations are invariant under the infinitesimal gauge transformations
\begin{equation}\label{eq:MCgaugetrafos}
 \delta \phi=-\sum_i \frac{(-1)^{i(i-1)/2}}{(i-1)!}\mu_i(\alpha,\phi,\cdots,\phi)~,
\end{equation}
where $\alpha$ is an element of $L$ of degree 0. The classical Maurer-Cartan equations $\dd \phi+\frac{1}{2}[\phi,\phi]=0$ appear as a special case for Lie 1-algebras. 

We see three ways in which $L_\infty$-algebras are concealed in our previous discussion. First of all, the semi-bracket on 1-forms introduced in section \ref{subsec:PoissonForms} yields a semi-strict Lie 2-algebra \cite{Baez:2008bu}. As Lie 2-algebras are 2-term $L_\infty$-algebras, the 1-form description on $S^3$, which transgresses to the loop space description yields an $L_\infty$-algebra. Second, it was noticed in \cite{Palmer:2012ya} that 3-Lie algebras are special cases of differential crossed modules. The category of the latter is equivalent to that of strict Lie 2-algebras \cite{Baez:2003aa}, and we arrive again at an $L_\infty$-algebra. Let us 
stress, however, that the 3-bracket of 3-Lie algebras cannot be interpreted as a ternary product $\mu_3$ in an $L_\infty$-algebra unless one gives up the grading \cite{IuliuLazaroiu:2009wz}.

Both the above appearances of $L_\infty$-algebras do not seem to provide any further insights into our discussion. We therefore turn to another interpretation advocated for $n$-Lie algebras in \cite{IuliuLazaroiu:2009wz}. There, it was shown that both the Nahm and the Basu-Harvey equations correspond to Maurer-Cartan equations in certain $n$-term $L_\infty$-algebras \cite{IuliuLazaroiu:2009wz}. We now briefly review these structures and demonstrate that the higher $\Pi_\omega$-Nahm equations also fit into this picture.

We start from the gauge covariant form of the Nahm equation with coupling constants put to 1,
\begin{equation}
 \dder{s}T^{m_1}+[A_s,T^{m_1}]= \eps^{m_1\ldots m_{p+1}}[T^{m_2},\ldots,T^{m_{j+1}}]~,
\end{equation}
where the $T^m$ are functions on $\CI$ with values in the $p$-Lie algebra $\CA$ and $A_s$ is the gauge potential with values in $\frg_\CA$, the Lie algebra of inner derivations of $\CA$. We choose $\CA$ to be the $p$-Lie algebra $\CA_{p+1}\cong\FR^{j}$ with generators $e_i$, $i=1,\ldots,p$ and bracket
\begin{equation}
 [e_{i_1},\cdots,e_{i_p}]=\eps_{i_1\cdots i_pj}e_j~.
\end{equation}
Note that the algebra of inner derivations of $\CA_{p+1}$ is $\frg_{\CA_{p+1}}\cong \sSO(p)$. Consider now the $L_\infty$-algebra $L=L_0+L_1+L_2$ with
\begin{equation}\label{eq:Lcomponents}
\begin{aligned}
 L_0&=\Omega^0(\CI)\otimes \frg_{\CA}~,\\
 L_1&=\left(\Omega^0(\CI)\otimes \CC\ell(\FR^{n+1})\otimes \CA\right)~\oplus~\left(\frg_{\CA}\otimes\Omega^1(\CI)\right)~,\\
 L_2&=\CC\ell(\FR^{n+1})\otimes \CA\otimes\Omega^1(\CI)~,
\end{aligned}
\end{equation}
where $\CC\ell(\FR^{n+1})$ is the Clifford algebra with $n+1$ generators $\gamma_\mu$, $\mu=1,\ldots,n+1$, satisfying $\gamma_\mu\gamma_\nu+\gamma_\nu\gamma_\mu=2\delta_{\mu\nu}$. An element $\phi$ of $L$ decomposes as 
\begin{equation}
 \phi=\underbrace{\lambda}_{L_0}+\underbrace{T^\mu \gamma_\mu+A\,\dd s}_{L_1}+\underbrace{S^\mu\gamma_\mu \dd s}_{L_2}~,
\end{equation}
where $\lambda\in L_0$, $T^\mu, S^\mu\in \Omega^0(\CI)$ and $A\in \Omega^0(\CI)\otimes\frg_{A_n}$. We define the following products:
\begin{equation}
\begin{aligned}
\mu_1(\phi)&=\dder{s}\lambda\,\dd s+ \left(\dder{s} T^\mu\right)\,\gamma_\mu\,\dd s~,\\
\mu_2(\phi_1,\phi_2)&=[\lambda_1,A_2]\,\dd s+[A_1,\lambda_2]\,\dd s+A_1\acton T^\mu_2\,\gamma_\mu\,\dd s-A_2\acton T^\mu_1\,\gamma_\mu\,\dd s\\
&~~~+\lambda_1\acton T^\mu_2\,\gamma_\mu-\lambda_2\acton T^\mu_1\,\gamma_\mu~,\\
\mu_p(\phi_1,\ldots,\phi_p)&=[T^{\mu_1}_1,\ldots,T^{\mu_n}_n]\,\eps_{\mu_1\cdots\mu_p\nu}\gamma_\nu\,\dd s~.
\end{aligned}
\end{equation}
For $p=2$, the two products defined above have to be added. The Maurer-Cartan equation \eqref{eq:MCeqs} for $\phi\in L$ with grading 1 correspond to the Nahm equation for $p=2$, the Basu-Harvey equation for $p=3$ and corresponding higher Nahm equations for $p>3$. Note that also the gauge transformations of the (higher) Nahm equations in gauge covariant form are given by the corresponding gauge transformations \eqref{eq:MCgaugetrafos} for an $\alpha\in L$ with grading 0.

It is now straightforward to generalize this observation to the $\Pi_\omega$-Nahm equations discussed in this paper. For this, consider again an $L_\infty$-algebra $L=L_0+L_1+L_2$, where in \eqref{eq:Lcomponents}, we replace $\CA$ with $\CC^\infty(M)$ and $\frg_{\CA}$ by the algebra of inner derivations of the $p$-Lie algebra induced on $\CC^\infty(M)$ by a Nambu-Poisson structure $\Pi_\omega$ on $M$. That is, $\frg_{\CA}$ is in general a sub-algebra of the algebra of volume preserving diffeomorphisms. The higher Nahm equations correspond then to the Maurer-Cartan equations for elements of $L$ with grading 1, and gauge transformations correspond to transformations \eqref{eq:MCgaugetrafos} for an $\alpha\in L$ with grading 0.

\section*{Acknowledgments}

We would like to thank Richard Szabo and Maciej Dunajski for discussions. SP and CS would also like to thank the Newton Institute, Cambridge, for hospitality, where a part of this work was completed. The work of DH was carried out at Durham University and supported by the Engineering and Physical Sciences Research Council (grant number EP/G038775). The work of SP and CS was supported by a Career Acceleration Fellowship from the UK Engineering and Physical Sciences Research Council.

\appendices

\subsection{Ends of manifolds and volume types of volume forms}\label{app:A}

In this appendix, we briefly review the notion of an end of a manifold and its volume.

Consider a topological space $M$ together with an ascending sequence $K_i\subset K_{i+1}$, $i\in \NN$, of compact subsets whose interiors cover $M$. Then $M$ has an end for every sequence $U_i\supset U_{i+1}$, where $U_i$ is a connected component of $M\backslash K_i$. For example, the real line $\FR$ has two ends, which are obtained from the sequence $K_i=[-i,i]$ with $U_i=(i,\infty)$ and $U'_i=(-\infty,-i)$. 

More generally, one defines an end of a manifold $M$ as an element of the inverse limit system $\{K, \mbox{components of}~M\backslash K\}$ indexed by compact subsets $K$ of $M$, cf.\ \cite{Greene:1979aa}. 

If $M$ is orientable and endowed with a volume form, we say that an end has a finite volume, if there is a compact set $K$ such that the volume of the component of $M\backslash K$ containing the end is finite. Otherwise, we say that the volume is infinite.



\end{document}